\documentclass{JHEP3}\newcommand{\format} {\JHEPformat}

\usepackage{amsmath}
\usepackage{epsfig}
\usepackage{latexsym}

\newcommand{\JHEPformat} {
\bibliographystyle{JHEP}
\newcommand{\maketitlepage} {}
\abstract{\theabstract}
\keywords{\thekeywords}
\preprint{\thepreprint}
}

\newcommand{\TITLE}[1] {\newcommand{\thetitle} {#1}\title{#1}}
\newcommand{\ABSTRACT}[1] {\newcommand{\theabstract} {#1}}

\newcommand{\ADDRESS}[1] {\newcommand{\theaddress} {#1}}
\newcommand{\DATE}[1] {\newcommand{\thedate} {#1}\date{#1}}
\newcommand{\KEYWORDS}[1] {\newcommand{\thekeywords} {#1}}
\newcommand{\PREPRINT}[1] {\newcommand{\thepreprint} {#1}}

\def\half{\frac{1}{2}}

\def\pa{\partial}

\def\rt{\rightarrow}

\newcommand{\be}{\begin{equation}}
\newcommand{\ee}{\end{equation}}
\newcommand{\bea}{\begin{eqnarray}}
\newcommand{\eea}{\end{eqnarray}}

\newcommand{\non}{\nonumber \\}
\newcommand{\CR}{\non\cr}

\def\a{\alpha}

\def\l{\lambda} 
 
\def\vf{\varphi}  

\def\ra{\rightarrow}

\TITLE{Non-critical supergravity ($d>1$) and holography}

\ADDRESS{School of Physics and Astronomy\\
  The Raymond and Beverly Sackler Faculty of Exact Sciences\\
  Tel Aviv University, Ramat Aviv, 69978, Israel
}

\author{Stanislav Kuperstein, Jacob Sonnenschein\\
School of Physics and Astronomy\\
The Raymond and Beverly Sackler Faculty of Exact Sciences\\
Tel Aviv University, Ramat Aviv, 69978, Israel\\
E-mail:
\email{kupers@post.tau.ac.il, cobi@post.tau.ac.il}
}

\ABSTRACT{In this paper we investigate the supergravity 
          equations of motion associated with non-critical ($d>1$)  type II
          string theories that incorporate RR forms. 
 Using a superpotential formalism we determine several classes
of solutions. In particular we find  analytic backgrounds 
with a structure of $AdS_{p+2}\times S^{d-p-2}$ and 
numerical solutions that asymptote a linear dilaton with 
a topology of $R^{1,d-3}\times R \times S^1$.
The SUGRA solutions we have found can serve as anti holographic descriptions of gauge theories in
a  large $N$ limit which is different than the one of the critical gauge/gravity duality.
 It is characterized by $N\rt \infty$ and $g_{YM}^2 N \sim 1$.
We have made the first steps in analyzing  the corresponding 
 gauge theory  properties  like  Wilson loops and the glue-ball spectra.

}

\DATE{March 2004}


\PREPRINT{TAUP-2767-04\\{\tt hep-th/0403254}}

\format

\begin{document}

\maketitlepage


\section{ Introduction}

Gauge/gravity holographic duality  has been proven to be a powerful
tool of studying strongly coupled gauge dynamics. 
However, the anti holographic picture of hadronic physics   
suffers from several limitations.

One such limitation, which is   shared by  all  confining  gauge theories  duals
 of critical
supergravity (SUGRA) backgrounds, 
 is the fact that generically 
their spectrum  includes on top of the ordinary hadronic states, Kaluza
Klein states of the same scale of mass as that of the  hadrons. 
So  far we do not know of a  mechanism   to  disentangle the desired hadronic
particles from the KK contamination.

One obvious way to overcome this problem is to abandon string theories in
critical dimensions and their corresponding SUGRA backgrounds, 
and instead study the holographic duality with non
critical string theories  and SUGRA models.
Needless to say that superstring theories of dimensions close to four are  
desired not only as candidates of the  theory of gauge dynamics  but also of 
the  theory of quantum gravity and grand unification.

Non-critical string theories in $d$ dimensional space-time are
characterized by having a Liuville mode on their  world-sheet which
combines together with additional $d-1$ coordinates. For $d\leq 2$
these theories have been shown to be consistent and it was believed
that one cannot pass the so called ``$c=1$ barrier'' (for bosonic
strings). However, in ~\cite{Kutasov:1990ua} ~\cite{Kutasov:1991pv} it was shown that non-critical
superstring theories for $d > 2$
that admit supersymmetry in space time are
consistent. An explicit construction of world sheet theory with 
${\cal N}=(2,2)$ supersymmetry which includes an ${\cal N}=2$  super Liuville was
proven to have in even dimensions, space time supersymmetry. 
The corresponding GSO projection eliminates the tachyons and by that
guarantees that  the non-critical superstring theories are consistent. 
It was then conjectured that the super Liuville theory is equivalent
to the CFT describing the Euclidean black hole which is nothing but
the CFT on a cigar.  The corresponding  ${\cal N}=2$ case was
discussed in ~\cite{Giveon:1999zm} ~\cite{Giveon:1999px}
~\cite{Giveon:1999tq}
and  in ~\cite{Hori:2001ax}.
In ~\cite{Giveon:1999zm} ~\cite{Giveon:1999px}
~\cite{Giveon:1999tq} it was  argued that certain non-critical superstring theories
follow a double scaling limit of critical ten dimensional superstring theories  on CY manifold with 
an isolated singularity. 

Non-critical type II SUGRA backgrounds are believed to be solutions 
of the equations of motion  associated with the low
energy effective action  of the type II non-critical superstring
theory. Recall that the equations of motion are in fact the 
$\beta$ function equations associated with the world sheet scale invariance
so that every solution of them is guaranteed to relate to scale invariant vacuum of the
corresponding string theory.

 The idea to extend the anti-holographic description of gauge theories
to non-critical SUGRA backgrounds was introduced first in ~\cite{Polyakov:1998ju}
where a proposal of a dual of pure YM in terms of a 5d non-critical 
gravity background was proposed.  Following this idea there were several attempts to 
find solutions of the non-critical effective action that are adequate as duals 
of gauge theories ~\cite{Klebanov:1998yy} ~\cite{Ferretti:1998xu}
~\cite{Ferretti:1999gj}
~\cite{Klebanov:1998yz} ~\cite{Garousi:1999fu} 
~\cite{Minahan:1999yr} ~\cite{Nekrasov:1999mn}
~\cite{Billo:1999nf} ~\cite{Armoni:1999fb} 
~\cite{Imamura:1999um} ~\cite{Ghoroku:1999bk}.
For instance in ~\cite{Armoni:1999fb} certain five dimensional non-critical type
0 backgrounds  were 
written down and where shown to admit  certain properties of  non-supersymmetric Yang Mills  
theory in four dimensions. 
In contrast to the $d<10$ theories, the non-critical supergravity with
$d \gg 10$ exhibits new features ~\cite{Maloney:2002rr} like dilaton potentials with
non-trivial minima at arbitrarily small cosmological constant and
$d$-dimensional string coupling.

From the study of the gravity duals of non-trivial gauge theories,
mainly the critical ones but also the non-critical ones, it has become clear that a radial 
direction of the bulk theory plays the role of the re-normalization scale of the related gauge theory.
That was the rational in ~\cite{Polyakov:1998ju} to look for a five dimensional
setup dual to the pure YM theory. For theories with supersymmetries
one has to incorporate 
also isometries of the background that will correspond to $R$
symmetries and global 
symmetries of the dual gauge theories. For example the classical 
$U(1)_R$ symmetry of the ${\cal N}=1$ SYM may follow from an isometry
of an additional $S^1$ part of the $d$ dimensional space time hence 
one would anticipate based on this argument, that a six dimensional 
space time may be the ``minimal'' dual of   ${\cal N}=1$ SYM.

The aim of this paper is to search for solutions of the equations of
motion of the low energy effective action associated with  non
critical type II string theories. In particular our goal is to look for
solutions that may serve as useful anti-holographic descriptions of
gauge theories.  
Rather than solving the second order equations of motion we transform the problem
to that of a set of BPS equations derived from a superpotential that relates to the potential 
of the system
\footnote{See i.g.
~\cite{Papadopoulos:2000gj} ~\cite {Borokhov:2002fm}
~\cite{Kuperstein:2003yt},
where the superpotential approach has been extensively used.}.  
The non-criticality nature is manifested by a term in the potential
that is proportional to the non-critical factor $c= 10-d$. In the formulation that we are using 
this contribution to the potential takes the form
\be
V_{\textrm{non-critical}}= \frac{c}{2} e^{2(n\lambda+k\nu-2\phi)}
\end{equation}
where $\phi$ is the dilaton, and the fields $e^{2\lambda}$ and $e^{2\nu}$ are the wrap factors of 
the world volume and compact transverse part of the metric respectively (see (\ref{stringmetric})).

We found several families of solutions.
With no RR fields we find the  solutions of the cylinder with linear dilaton, the cigar its T-dual
the trumpet solution and certain generalizations of them
\footnote{See ~\cite{Giveon:1994fu} for a review of T-duality.}.
One class of solutions  with RR fields 
are of the form $AdS_{n+1}\times S^k$ with $k\neq 1$,  $n+1\neq k$ and 
where the total dimension is $d= n+k+1$.
Close relatives of these solutions are near extremal solutions based on these  $AdS_{n+1}\times S^k$
solution. We re-derive using the superpotential approach the charged 
black hole solution of ~\cite{Berkovits:2001tg}
including its near horizon $AdS_2$ solution as well as their T-dual solution.
As mentioned above backgrounds that incorporate a cylindrical geometry 
and linear dilaton ~\cite{Aharony:1998ub} has special interest both
from the point of view of the world sheet description of non-critical superstring theory as well as
from the point of view of backgrounds that correspond to brane configurations associated with 
interesting gauge theories. We therefore put a special emphasis in looking for solutions that admit this
geometry. Indeed we found solutions that asymptote $R^{1,d-3}\times R
\times S^1$ and incorporate RR fields. 
We present the corresponding numerical solutions in the full range of the radial direction 
 and the approximated analytic behavior both
asymptotically as well as close to the origin of the radial direction. 
In this paper we will not discuss the exact superstring solutions
in non-trivial non-critical space-times with NS-NS and no RR 
fields based on superconformal world-sheet
supersymmetry (see ~\cite{Kounnas:1994py} for a review).

A generic feature of the solution is that due to the non-criticality factor the 
scalar curvature of various background is not small but rather is proportional to $c= 10-d$.
Therefore neglecting higher curvature terms is not really justified. The conjecture is
that for special solutions like the  $AdS_{n+1}\times S^k$, the structure of the background would not be 
modified  by the higher order curvature contributions but rather only the corresponding radii will be 
corrected. 

For the classes of solutions we have discovered, we analyze the holographic boundary field theories.
We substantiate the idea that the boundary field theory is a gauge theory by proving that the 
Bekenstein Hawking entropy of the SUGRA background scales like $N^2$, where $N$ is the flux of the 
RR form.  We show that the anti-holographic SUGRA picture of the gauge field corresponds to a 
large N limit where
\be
N\rt \infty \qquad g_{YM}^2 N \sim 1,
\end{equation}  
 which is between perturbative  large $N$ limit and the one used in the gauge/gravity duality.
We briefly discuss the Wilson loop and the glue-ball spectrum for duals of pure YM in 3 and 4 dimensions
that follow from the large temperature limit of the near extremal  $AdS_{n+1}\times S^k$ solutions.
We also address the these properties for the theories, which may be viewed as the RR deformation
of the cigar solution.

The paper is organized as follows: in the next section we present the
general  setting, namely the non-critical low energy effective action,
the corresponding equations of motion and the superpotential approach
that leads to BPS equations compatible with the equations of motion.
Section \ref{Dp} is a warmup exercise where we find the $D_p$ solutions by
switching off the non-criticality term. Section \ref{Q0} is devoted to
non-critical backgrounds that do not include NS or RR forms. In subsection
\ref{Q0k0} we present solutions of backgrounds with no transverse directions,
In subsection \ref{k=1} we derive the cylinder solution, the cigar solution
and certain generalization of it. The next subsection describes the
cigar and trumpet solutions as T-duals and in  the last subsection
we find a special solution in four dimensional compact transverse
space.
Section \ref{Conformal} is devoted to the  conformal $AdS_{n+1}\times S^k$ backgrounds
both as solutions of the equations of motion as well as solutions of
the BPS equations. In subsection \ref{AdSBH} we describe the AdS black hole
associated with the   $AdS_{n+1}\times S^k$ backgrounds.
In Section \ref{RRdef2dimBH}  we describe two dimensional solutions that incorporate RR
fields. In particular we derive the charged black hole solution of
~\cite{Berkovits:2001tg}, take the near horizon limit that leads to  the $AdS_2$
solution, and construct the T-dual of the charge black hole solution.
Section \ref{RRdef} is devoted to solutions that include RR forms in the
cylindrical geometry or differently stated RR deformations of the
cigar/trumpet solutions.  We describe numerically three families of
solutions. We derive approximate expressions for these backgrounds
both asymptotically as well as close to the origin.
A brief discussion of backgrounds that include NS three form and no RR
forms is presented in Section \ref{QNS}. In Section \ref{holography} we explore the 
properties of the holographic dual gauge theories associated
with the background solutions we had found. We first show that the
entropy of the non-critical SUGRA backgrounds admit entropy that
scales like $N^2$. We then show in \ref{AnovelLimit} that the non-critical duality
in fact holds for a novel large N limit where $g_{YM}^2 N\sim 1 $. 
We then describe in 
\ref{TheGaugeTheories} the gauge theories which are the cousins of the   
$AdS_{n+1}\times S^k$ backgrounds. In \ref{AdSBHandDualGT} we analyze
the dual gauge
theory of the AdS black hole solutions in particular the Wilson loops
and the glue-ball spectra are addressed. The duals of the RR deformed
cylinder SUGRA backgrounds are described in subsection 
\ref{DualRRdef}.
To the benefit of the reader we include several appendices with
certain explicit computations of the quantum mechanical effective
action, The BPS equations both in the string an Einstein frames, 
certain potential - superpotential relations, the derivation of the
cigar solution in the Einstein frame and the solution of
~\cite{Berkovits:2001tg} from
the equations of motion.

\section{General setting}

The metric in the \emph{string} frame is taken to depend only on  the
radial coordinate $\tau$.
It takes the form  
\begin{equation}       \label{stringmetric}
l_s^{-2}ds^2 = d\tau^2 +  e^{2\lambda(\tau)} dx_{\|}^2 +e^{2\nu(\tau)} d\Omega_k^2
\end{equation}
where $dx^2_{\|}$ is $n$ dimensional flat metric, and
$d\Omega_k^2$ is a $k$ dimensional sphere. 

The bosonic part of the  
 non-critical  SUGRA action in $d$ dimensions takes the form
 
\bea       \label{eq:TheAction}
S &=& \int d^{n+k+1} x \sqrt{G} e^{-2 \phi} 
               \left( R+ 4(\pa\phi)^2 + \frac{c}{\alpha^\prime} \right)\CR 
 && - \frac{e^{-2 \phi}}{2}  \int H_{(3)} \wedge \star H_{(3)} 
    - \sum_{p} \half   \int F_{(p+2)} \wedge \star F_{(p+2)} , 
\eea
where 

\be
\frac{c}{\alpha^\prime}=\frac{10-d}{\alpha^\prime}
\end{equation}
is the non-criticality central charge term
\footnote{Throughout this paper
          $c$ is a dimensionless parameter ($c=10-d$). Also all
          the coordinates on the r.h.s. of
          (\ref{stringmetric}) are dimensionless in accordance
          with the $l_s^{-2}$ factor on the l.h.s.}.     
The latter term corresponds to the deviation  of the dilaton beta function
at non-critical space-time dimension from the one at critical dimension.
 $F_{p+2}$ is a RR form
that corresponds to a $D_p$ brane with $n=p+1$ dimensional world volume, and
$ H_{(3)}$ is the NS three form. 
Throughout this paper we will investigate solutions with only one non
trivial RR form.

Upon substituting the metric (\ref{stringmetric}) into  the action and performing
the integration  one finds ~\cite{Klebanov:1998yy} 
(see Appendix \ref{ThaActionDerivation} for the derivation): 
\bea \label{actionrho}
S = l_s^{-2}\int d\rho  \left( \left[-n(\lambda')^2 -k(\nu')^2 +(\varphi')^2
+ c e^{-2 \varphi} + (k-1)k e^{-2\nu-2 \varphi} \right]\right)+ S_{RR} + S_{NS},
\eea
where $d\tau= -e^{-\varphi}d\rho$, $(A)'=\pa_\rho A$ and 
\be
\varphi=2\phi-n\lambda-k\nu
\end{equation}  
is the
``shifted'' dilaton.

Assuming that the RR form also depends only on the radial direction, namely, 
$F=\pa_\tau A dx^0\wedge...dx^p\wedge d\tau$, the RR 
part of the action reads 
\be\label{SRR}
S_{RR}= -\int d\rho \left(\frac{1}{4} e^{-n\lambda + k\nu-\varphi} (A')^2 \right ) =
-  Q^2 \int d \rho e^{n \lambda -k \nu - \varphi},
\end{equation}
where we made the substitution  $A'= 2Qe^{n\lambda-k\nu-\varphi}$  which is 
the solution of the equation of motion of  $A''-A'(n\lambda'-k\nu'-\phi')=0$ .
In fact the assumption was  that the Hodge dual of the 
RR form is proportional to the volume form $\omega_k$ of the transversal $S^k$,
namely $\star F_{(p+2)} = N \omega_k$, where $p=n-1$ and $N$ is the
 number of the $Dp$-branes. 
$Q$ is equal $N$ up to some numerical factor coming, in
particular, from the integration over the $S^k$ part of the metric.
Alternatively we can assume that $F_{(p+2)} = N \omega_k$. 
This way we will end up with a stack of $D{p}$, where this time
$p=k-2$.
 For instance, for $k=1$ and $n=4$   the corresponding form is  $F_{(5)}$ so
we take its Hodge dual  $F_{(1)} \sim d \theta$ 
(the $0$-form potential $A_{(0)}$ depends linearly on $\theta$)
where $\theta$ is the coordinate on the $S^1$. Alternatively, this can
be viewed as 
 $F_{(5)}=\star_{6} F_{(1)} \sim \star d \theta$ ($A_{(4)}$ is a function
of $\tau$). In the former case  this is interpreted as
 a stack of $D3$-branes and
in the later case we have "magnetic" duals, namely $D(-1)$-branes. In
both cases the form contribution (\ref{SRR}) to the action is
the same.

Similarly, for $k=3$ the NS 3-form is given by
 $H_{(3)}=N \omega_3$ with $N$ denoting the number
of the NS5 branes.
Under these assumptions upon substituting the solution of the equation of motion for the 
NS forms, the NS action reads 
\be \label{SrrSns}
S_{NS}= - Q^2 \int d \rho  e^{- 2k \nu- 2 \varphi}.
\end{equation}
The second order equations of motion are:
\bea   \label{EMrho}
&& \pa_{\rho}^2\lambda  - \frac{1}{2}Q_{RR}^2  e^{n\lambda-k\nu-\varphi}  = 0, \CR
&& \pa_{\rho}^2\nu -(k-1) e^{-2\nu-2\varphi}+ \frac{1}{2} Q_{RR}^2 e^{n\lambda-k\nu-\varphi} 
   + Q_{NS}^2 e^{-2k \nu -2 \varphi} = 0,  \CR
&& \pa_{\rho}^2\varphi+ (c + (k-1)k e^{-2\nu})e^{-2\varphi} 
- \frac{1}{2} Q_{RR}^2 e^{n\lambda-k\nu-\varphi}
 - Q_{NS}^2 e^{-2k \nu -2 \varphi} = 0. 
\eea
In terms of $\lambda$,$\nu$ and the dilaton $\phi$  it takes the form
\bea  \label{EMrhophi}
&&   \pa_{\rho}^2\lambda - \frac{1}{2} Q_{RR}^2 e^{2(n\lambda-\phi)}   = 0,     \CR
&&   \pa_{\rho}^2\nu -(k-1) e^{2(n\lambda+(k-1)\nu -2\phi)} 
+ \frac{1}{2} Q_{RR}^2 e^{2(n\lambda-\phi)} + Q_{NS}^2 e^{2(n\lambda-2 \phi)}  = 0,  \CR
&&    \pa_{\rho}^2\phi+ \frac{c}{2} e^{2(n\lambda+k\nu-2\phi)} 
      - \frac{(n+1-k)}{4} Q_{RR}^2 e^{2(n\lambda-\phi)} + \frac{(k-1)}{2} Q_{NS}^2 e^{2(n\lambda-2\phi)}
       = 0. 
\eea
Again the NS term is relevant only for $k=3$.
Solving these equations one also has to imply the zero-energy constraint, 
in other words the Hamiltonian derived from the 1d action
(\ref{actionrho}) must be set to zero. 
This constraint which is in fact the equation of motion associated with  $g_{\tau\tau} $
takes the following form
\be
n(\pa_\tau\lambda)^2 +k(\pa_\tau\nu)^2 -(\pa_\tau\varphi)^2
+ c  + (k-1)k e^{-2\nu}- Q_{RR}^2 e^{n\lambda-k\nu+\varphi}- 
Q_{NS}^2 e^{-2k \nu } =0.
\end{equation}


\subsection{ The superpotential and BPS equations}

It is well known that for supersymmetric backgrounds one can avoid the
hurdle of solving  second
order differential equations and instead solve first order BPS equations. 
The solutions of the latter equations are also solutions of the equations of motion.   

Consider the following general form of a background action
\be
S = \int d\rho \left(-\half G_{ab}{f^a}'{f^b}' -V(f)\right)
\end{equation}
The corresponding equations of motion are given by
\bea\label{seqom}
0 &=&
(G_{ab} {f^b}')' - \half \pa_a G_{bc} {f^b}' {f^c}'
 -\pa_a V
\non
 &=&
 G_{ab} {f^b}'' + (\pa_c G_{ab} - \half\pa_a G_{bc}){f^b}' {f^c}'
 -\pa_a V
\non
 &=&
  G_{ab}({f^b}'' + \Gamma^b_{cd} {f^c}' {f^d}' - \pa^b V).
\non
\eea
and the zero-energy constraint reads $- \half G_{ab}{f^a}'{f^b}' + V(f)=0$. 

If the potential is  related in the following way  to a superpotential $W$
\be
V =\frac{1}{8} G^{ab}\pa_aW\pa_bW, 
\end{equation}
then the BPS equations are 
\be
{f^a}' = \half G^{ab}\pa_bW.
\end{equation}
It is straightforward to check that the BPS equations are compatible with the equations
of motion (see Appendix \ref{FromBPSToEquations}) and with the zero energy condition.
Applying this structure to the action of (\ref{actionrho}) we have that 
\be
G_{\lambda\lambda}=2n \qquad G_{\nu\nu}=2k\qquad G_{\varphi\varphi}=-2
\end{equation}
and
\be
V = {Q^2} e^{n\lambda-k\nu-\varphi}- (c + (k-1)k e^{-2\nu}) e^{-2\varphi} .
\end{equation}
and therefore the relation between the potential and the superpotential reads
\be          \label{potsup}
\frac{1}{n} (\pa_\lambda W)^2  + \frac{1}{k} (\pa_\nu W)^2 -  (\pa_\varphi W)^2 = 16 V
\end{equation} 
and the BPS  equations are 
\be
\lambda' = \frac{1}{4n}\pa_\lambda W,\qquad
 \nu' = \frac{1}{4k}\pa_\nu W,\qquad
\varphi' =  -\frac{1}{4}\pa_\varphi W.
\end{equation}

Another useful parameterization of the superpotential approach is the following. Consider
the action that takes the following form 
\be\label{actionE}
S = \int du\,
   e^{4A}\left(3 (A')^2- \half G_{ab}{f^a}'{f^b}' -V(f)\right),
\end{equation}
where now $\prime$ denotes derivative with respect to $u$.
In this formulation we look for a   corresponding $W$ which is related
to the potential as follows 
\be
V =\frac{1}{8} G^{ab}\pa_aW\pa_bW - \frac{1}{3} W^2
\end{equation}
and the BPS equations are
\be
{f^a}' = \half G^{ab}\pa_bW \qquad A'= - \frac{1}{3} W(f).
\end{equation}
This formulation is adequate for the action in the Einstein frame.
The reformulation of the whole
background in the Einstein frame and the relation between the fields 
in the string frame to those in the Einstein frame are presented 
in Appendix \ref{EinsteinFrame}.


 \section{$Dp$ brane solutions in critical dimension ($c=0,\  d=10$)}

\label{Dp}
  
Let us now as a warmup exercise apply the formalism of the
superpotential and the BPS first order differential  equations for the
case of critical superstring, namely for ten dimensional space time.
For the supergravity background with a RR form and no NS form, 
the superpotential equation has a simple solution:

\be
W= 2 \sqrt{2} Q e^{\frac{1}{2} (n \lambda - k \nu -\varphi)} - 4 k e^{-\nu-\varphi}.
\end{equation}
Let us write the BPS equations in terms of a radial coordinate $r$
defined by 

\be
e^{-\varphi} d \rho = e^{\nu}\frac{d r}{r}. 
\end{equation}
We get:
\bea   
&& r \frac{d \lambda }{d r}  = 
 \frac{\sqrt{2}}{4} Q e^{\frac{1}{2} (n \lambda - (k-2) \nu +\varphi)}  \CR
&& r \frac{d \nu }{d r}  =
 - \frac{\sqrt{2}}{4} Q e^{\frac{1}{2} (n \lambda - (k-2) \nu +\varphi)} + 1 \CR
&& r \frac{d \varphi }{d r}  =
  \frac{\sqrt{2}}{4} Q e^{\frac{1}{2} (n \lambda - (k-2) \nu +\varphi)} - k.
\eea
Due to the fact that the exponential factor in all these equations
is the same it is useful to 
define $\a=n \lambda - (k-2) \nu +\varphi$ which obeys the following equation

\be
 r \frac{d \a}{d r}  = 2 \sqrt{2} Q e^{\a/2} - 2(k-1),
\end{equation}
which is solved by:

\be
e^{-\a/2} = g_s^{-1} r^{k-1} + \frac{\sqrt{2} Q}{k-1}.
\end{equation}
Plugging this into the equations for $\lambda$ and $\nu$ we find the
expected result of the $Dp$-brane metric $ds^2 =e^{2 \lambda}
dx_\mu^2+ r^{-2} e^{2 \nu}(dr^2 + r^2 d\Omega^2)$, where:

\be
e^{2 \lambda} = \left( 1 + \frac{ \frac{\sqrt{2}}{7-p} g_s Q}{r^{7-p}}\right)^{-1/2}
 \qquad \textrm{and} \qquad
e^{2 \nu} = r^2 \left( 1 + \frac{ \frac{\sqrt{2}}{7-p} g_s Q}{r^{7-p}} \right)^{1/2},
\end{equation}
where we have replaced $k-1=(10-n-1)-1= 7-p$. The string coupling
constant $g_s$ is a free parameter of the solution and is  
set to be equal to an asymptotic value of the
dilaton:  
$e^\phi\vert_{r \to \infty} = g_s$.

Using this result we can calibrate for the critical case the value of 
$Q$ by comparing to the metric of the $Dp$ brane. The result is 
\be
Q= \frac{7-p}{\sqrt{2}}2^{7-2p}\pi^{\frac{2-3p}{2}} \Gamma \left( \frac{7-p}{2} \right)  N.
\end{equation}

 
 \section{Solutions with zero RR and NS-NS charges ($Q=0$)}

\label{Q0}

With no RR or NS-NS forms the potential include the non-critical term 
as well as a curvature term from 
the $S_k$ part of the metric. There is  no contribution form the
curvature for $k=0$ and $k=1$ as can 
be seen from 
\be
V =  - k(k-1) e^{-2 \nu - 2 \varphi } - c e^{-2 \varphi}.
\end{equation}
The corresponding superpotential still has to solve (\ref{potsup})
which can be simplified due to the
$e^{-2 \varphi}$ factor which is common to both  terms of the
potential.
This obviously calls for an ansatz of the form
\be
W(\l,\vf,\nu) = 4e^{-\varphi}w(\l,\nu) 
\end{equation}
for which (\ref{potsup}) takes the form
\be\label{potsupk}
 \frac{1}{n} (\pa_\lambda w)^2  + \frac{1}{k} (\pa_\nu w)^2 - w^2 =
  - k(k-1) e^{-2 \nu } - c 
\end{equation}
Due to the absence of the curvature part, the cases of no transverse $S_k$ and of  $S_1$ are 
special so we start with analyzing them first.


\subsection{ No transverse sphere ($k=0$)}

\label{Q0k0}

In this case the metric is
\be
l_s^{-2} ds^2 = e^{2 \lambda} dx^2_{\|} + d \tau^2
\end{equation}
The equation (\ref{potsupk}) for the superpotential is solved by 
\be
w(\lambda) = \pm  \sqrt{c}
 \quad \textrm{or} \quad w(\lambda) = \pm \sqrt{c}\cosh (\sqrt{n} \lambda).
\end{equation}
For the first type of solution we end up with the following BPS equations 
\be
\lambda' = 0 
\qquad
 \varphi' = \pm \sqrt{c}  e^{-\varphi}.
\end{equation}
With no loss of generality we can set $\l=0$ so the background
includes  a $(n+1)$ dimensional  flat Minkowskian metric
\be
l_s^{-2} ds^2 = -dt^2 + \ldots + dx_{n-1}^2 + d\tau^2
\end{equation}
and a linear dilaton
\be         \label{eq:LinearDilaton}
e^\varphi=\pm \sqrt{c} \rho\ \  \ra  \qquad \varphi=\pm \sqrt{c} \tau
\ \ 
  \ra \qquad \phi =\pm \frac{\sqrt{c}}{2}  \tau. 
\end{equation}
 Note  that in 10d the dilaton becomes constant
 and  the ordinary ten dimensional  flat Minkowski solution is retrieved.
In fact the background with the linear dilaton corresponds to an exact 2d conformal theory on
the world-sheet ~\cite{Myers:1987fv}.

For the second type of superpotential, namely,  $w(\lambda) = \pm
\sqrt{c}\cosh (\sqrt{n} \lambda)$ the BPS equations take the following form
\be      \label{BPSII}
\l'=\sqrt{\frac{c}{n}} e^{- \varphi} \sinh(\sqrt{n}\l), 
\qquad \vf'= \sqrt{c}e^{- \varphi} \cosh(\sqrt{n}\l). 
\end{equation}
By transforming to derivatives with respect to $\tau$ these equations can be solved analytically. 
Instead we will now analyze  the equations of motion and
then come back to the  solution of the BPS equations.

Indeed 
for the case of  vanishing  RR and  NS-NS forms at $k=0$ (and $k=1$) 
the equations of motion (\ref{EMrho}) are simple enough to be solved directly. 
The equations of motion 
\be
\l''=0, \qquad \vf'' + ce^{-2\varphi}= 0
\end{equation}
are solved by 
\be           \label{soleqom}
\lambda' = b ,
\qquad
 \varphi' = \sqrt{c} \left(a^2 + e^{-2\varphi}\right)^{1/2}.
\end{equation}
Plugging this into the zero-energy condition we find:
\be
n b^2 = c a^2.
\end{equation}
For $a=b=0$ the equation for $\varphi$ is solved by $e^\varphi=\sqrt{c}  \rho$ or 
 $\varphi=\sqrt{c} \tau$ so that  the string coupling is
$e^{2 \phi} \sim e^{\sqrt{c} \tau} $.
This is obviously the linear dilaton solution associated with the
superpotential $w=\sqrt{c}$ which was discussed above.

\FIGURE[b]{
 \label{DilatonAGH}
\centerline{\input{DilatonAGH.pstex_t}}
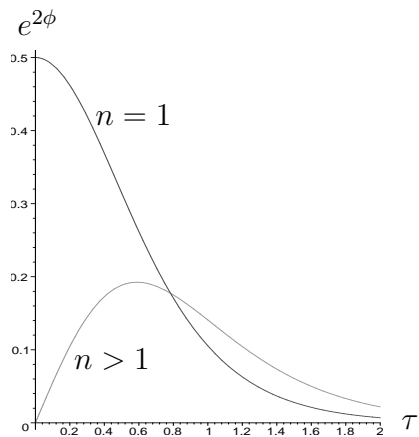
\caption{The string coupling vs. the radial coordinate $\tau$ for $a=1$ and $n=1,4$. 
         According to (\ref{eq:DilatonAGH}) the maximum value of
         $e^\phi$ is fixed by $a$ and for large enough $a$ the 
         string coupling is small everywhere.
} 
}

For $a \neq 0$  the solution of the equations reads 

\be
\lambda = - \sqrt{\frac{c}{n}} \rho, 
\quad  \quad
e^\varphi = \frac{1}{a} \sinh( \sqrt{c} a \rho)
\end{equation}
or in terms of the $\tau$ coordinate
\be
e^\lambda = \left[ \tanh \left( \half \sqrt{c} \tau \right)   \right]^{1/\sqrt{n}}
\quad \textrm{where} \quad
e^{\sqrt{c} \tau} = \tanh \left( \half \sqrt{c} \rho\right). 
\end{equation}
Now let us come back to the second type of superpotential discussed above. 
Plugging this solution into 
(\ref{soleqom}) it is easy to verify that indeed equations
(\ref{BPSII}) 
are identical  to  (\ref{soleqom}).
Thus we have established a one to one map between the solutions of the
BPS equations and the solutions of the 
equations of motion. 
The dilaton derived from this solution is given by:

\be   \label{eq:DilatonAGH}
e^{2 \phi} = \frac{1}{a} \frac{ \left[ \tanh \left( \half \sqrt{c} \tau \right) \right]^{\sqrt{n}}}
                         {\sinh\left( \sqrt{c} \tau \right) }
\end{equation} 
and we see that for large enough $a$ we have a small string coupling
at any $\tau$ (see Fig. \ref{DilatonAGH}). 
For $\tau \to \infty$ the solution reduces to the
linear dilaton background with the flat Minkowskian metric discussed above.
The scalar curvature associated with this metric is given by 
\be     \label{eq:CurvAGH}
l_s^2 \mathcal{R}  = -\frac{c(n+1)}{\sinh^2 \left( \sqrt{c} \tau \right)}
         \left( \frac{2 \sqrt{n}}{n+1} \cosh \left( \sqrt{c} \tau \right) -1 \right).
\end{equation}
For $n=1$ the metric is regular everywhere and for $n> 1$ there is a naked singularity at 
$\tau =0$ (see Fig.\ref{CurvAGH}).

\FIGURE[b]{
 \label{CurvAGH}
 \centerline{\input{CurvAGHn1.pstex_t} \input{CurvAGHn4.pstex_t}}
 \caption{The curvature (\ref{eq:CurvAGH}) vs. $\tau$. For $n=1$
 there is no singularity, while for $n>1$ the metric is singular at
 $\tau \to 0$. 
} 
}

This solution was originally derived in ~\cite{Alvarez:2000it} in the context of bosonic
strings. It was further shown that in the linear approximation the closed
strings tachyon equation of motion has an exact "kink" solution,
which interpolates between different expectation values.


\subsection{$k=1$,  the cylinder, the cigar and beyond}
\label{k=1}

In this case the compact transverse space is an $S^1$. Let's start
again with the
equations of motion
\be
\pa^2_\rho \lambda =0, \quad  
\pa^2_\rho \nu =0, \quad  
\pa^2_\rho \varphi = -c e^{-2 \varphi}. 
\end{equation}
So that:

\be
\pa_\rho \lambda = b_\lambda \rho, \quad  
\pa_\rho \nu = b_\nu \rho, \quad  
\pa_\rho \varphi = c^\frac{1}{2} \left( a^2 + e^{-2 \varphi} \right)^{1/2}. 
\end{equation}
The zero energy condition implies  $n b_\lambda^2 + b_\nu^2 = ca^2$ so it is convenient to 
parameterize the solution as follows
\be
b_\lambda = \sqrt{ \frac{c}{n}} a \cos \beta
\quad  \textrm{and} \quad
b_\nu = \sqrt{c} a \sin \beta.
\end{equation}
For $a=0$  $\lambda$  and $\nu$ are constants.
Again with no loss of generality we take them to be $\lambda=\nu=0$ so that we get a geometry
of a \emph{cylinder}: 
\be\label{cylinder}
ds^2 = -dt^2 + \ldots + dx_{n-1}^2 + d\tau^2 + d\theta^2
\end{equation}
with a linear dilaton
\be 
\phi =\half\sqrt{ c} \tau. 
\end{equation}
As we have already mentioned this solution corresponds to an exact 2d conformal theory on
the world-sheet ~\cite{Myers:1987fv} (now one of the coordinates is compact).
\FIGURE[b]{
 \label{CigarDilCurv}
 \centerline{\input{CigarDilCurv.pstex_t}}
 \caption{The curvature (\ref{eq:CigarCurv}) and the dilaton (\ref{eq:dilatoncigar}) of the
   "cigar" solution for $a=1$ and $c=4$. 
} 
}
A more interesting background is found for 
 $a \neq 0$ where 
we have $e^{-c^\half \tau} = \tanh \left( \half \sqrt{c} a \rho \right) $ and hence

\be
e^\lambda = \left[ \coth \left( \half \sqrt{c} \tau \right) \right]^{\frac{\cos \beta}{\sqrt{n}}}
\quad
e^\nu = \left[ \coth \left( \half \sqrt{c} \tau \right) \right]^{\sin \beta}.
\end{equation}
Substituting this into the expression for the curvature and requiring regularity
at $\tau=0$ we obtain the following equation for $\beta$:

\be      \label{betan}
\sqrt{n} \cos \beta + \sin \beta = -1,
\end{equation}
which is solved by by two types of solutions. One solution  is simply
$\sin \beta = -1$.
This implies $\lambda=0$ and
$e^\nu =  \tanh \left( \half \sqrt{c} \tau \right)$,  which is nothing but the famous
{\it cigar} background, or more prices $R^{1,p}\times\ cigar$, namely
\be     \label{eq:metriccigar} 
ds^2 = -dt^2 + \ldots + dx_{n-1}^2 + d\tau^2 + \tanh^2 \left( \half \sqrt{c} \tau \right) d\theta^2
\end{equation}
with a  dilaton of the form 

\be  \label{eq:dilatoncigar} 
e^{2 \phi} = \frac{1}{2 a} \frac{1}{ \cosh^2 \left(\half \sqrt{c} \tau \right)}.
\end{equation}
The radius of the compact coordinate in the metric
(\ref{eq:metriccigar}) is equal to $R_{\theta}=\frac{2}{\sqrt{c}}$. It
is fixed by requiring the space to be regular at $\tau=0$.

The scalar curvature of this ``cigar'' background is (see Fig. (\ref{CigarDilCurv})):

\be            \label{eq:CigarCurv}
l_s^2 \mathcal{R} = - \frac{c}{ \cosh^2 \left(\half \sqrt{c} \tau \right)}.
\end{equation}
The cigar background like the cylinder one corresponds to an exact
string solution ~\cite{Callan:1991at} (see also ~\cite{Murthy:2003es}
a recent discussion of the topic). 
\FIGURE[b]{
 \label{NewCigar}
 \centerline{\input{NewCigar.pstex_t}}
 \caption{The warp factors $e^\lambda$ and $e^\nu$ (\ref{eq:k1})
   vs. $\tau$ for $c=4$.  
} 
}

The other solution of eqn. (\ref{betan}) is 
\be
\cos \beta = - \frac{2 \sqrt{n}}{n+1}, \quad \sin \beta =\frac{n-1}{n+1}.
\end{equation}
The corresponding  components of the metric read (see Fig. \ref{NewCigar})
\be  \label{eq:k1}
e^\lambda = \left[ \tanh \left( \half \sqrt{c} \tau \right) \right]^{\frac{2}{n+1}}
\quad
e^\nu = \left[ \tanh \left( \half \sqrt{c} \tau \right) \right]^{-\frac{n-1}{n+1}}.
\end{equation}  
Remarkably, the dilaton and the curvature of this solution
are precisely like in the cigar background.  The two backgrounds,
however, are not related through a coordinate transformation for $n>0$, as can
be seen from the non-trivial warp factor $e^\lambda$. 
As a consistency check one may easily verify that for $n=0$ the solution reduces to the cigar
background without the $R_{n-1,1}$ part in the metric.
In  contrast to the cigar solution the metric given by (\ref{eq:k1})
has a horizon at $\tau=0$. Moreover, this point cannot anymore be thought of
as the tip of the cigar since (at least for $n>1$) the warp factor
$e^\nu$ diverges at $\tau=0$.

It will be interesting to see whether this configuration with the non-trivial
warp factor $e^{\lambda(\tau)}$ might appear as an exact world-sheet solution similar to the
cigar background.

Before moving on to higher dimensional transverse spaces, let us
briefly remark about the superpotential approach to the $k=1$ case.  
A superpotential that corresponds to $V=ce^{-2\varphi}$ for the case of a non-trivial $\nu$ is
\be
W =-  4 \sqrt{c}e^{-\varphi} \cosh(\nu).
\end{equation}
The corresponding BPS equations are 
\bea      \label{eq:BPScigar}
\pa_\rho\varphi &=& - \frac{1}{4} \pa_{\varphi}W =  -\sqrt{c}e^{-\varphi} \cosh(\nu) \CR
\pa_\rho\nu &=& \frac{1}{4} \pa_{\nu}W = -\sqrt{c}e^{-\varphi} \sinh(\nu). \CR
\eea
To solve the corresponding BPS equations it is convenient to rewrite these equations
in terms of derivatives with respect to $\tau$. The second equation then reads

\be       \label{eq:BPScigarnu}
\pa_\tau\nu =  \frac{1}{4} e^\varphi  \pa_{\nu}W = - \sqrt{c} \sinh(\nu),
\end{equation}
which admits a solution of the form 
\be
e^\nu =  \tanh \left( \half \sqrt{c} \tau \right). 
\end{equation}
Plugging this back into the first equation in (\ref{eq:BPScigar}) one finds the dilaton solution
of (\ref{eq:dilatoncigar}).
This combines with $e^{2\lambda}=1$ 
is obviously the cigar solution. For completeness we present in
Appendix \ref{TheCigarEinsteinframe} the derivation of the cigar solution in the Einstein frame
using the formulation of (\ref{actionE}).


\subsection { The cigar and trumpet solutions as  T-duals}

\FIGURE[b]{
 \label{Bell}
 \centerline{\begin{picture}(0,0)%
\includegraphics{Bell.pstex}%
\end{picture}%
\setlength{\unitlength}{2447sp}%
\begingroup\makeatletter\ifx\SetFigFont\undefined%
\gdef\SetFigFont#1#2#3#4#5{%
  \reset@font\fontsize{#1}{#2pt}%
  \fontfamily{#3}\fontseries{#4}\fontshape{#5}%
  \selectfont}%
\fi\endgroup%
\begin{picture}(4212,4213)(139,-3587)
\put(3976,-1336){\makebox(0,0)[b]{\smash{{\SetFigFont{12}{14.4}{\rmdefault}{\mddefault}{\updefault}$e^\nu$}}}}
\put(4351,-1786){\makebox(0,0)[b]{\smash{{\SetFigFont{12}{14.4}{\rmdefault}{\mddefault}{\updefault}$\tau$}}}}
\end{picture}%
}
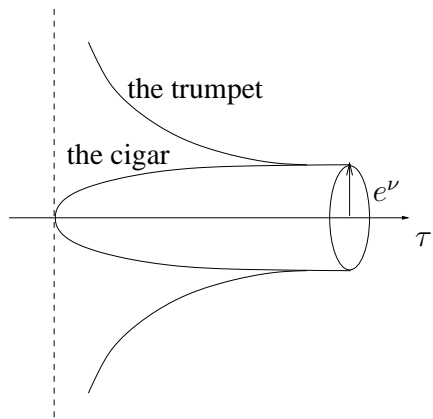
 \caption{The "cigar" and the "trumpet" geometry. In the former the
   factor $e^\nu$ vanishes at $\tau=0$ (the tip), while in the later
   $e^\nu \to \infty$ and the curvature diverges.
} 
}

In the last subsection we have discussed solutions characterized by a
compact $S_1$ transverse space.
This naturally calls for the implementation of T-duality to generate 
new solutions of the equations of motion. In the present context 
T-duality acts on $e^{\nu}$ and on $e^\phi$ as follows
\be
e^{2\nu}\ \ \rt e^{- 2\nu} \qquad e^{2\phi} \ \ \rt  {e^{2\phi-2\nu}}
\end{equation}
where we still use $\alpha'=1$. The cylinder solution given in
(\ref{cylinder}) is unaltered by T duality. 
For a constant radius different from 1 the transformation is also obvious.

The more interesting cases are  of course the cigar solution and its generalizations.
Replacing $\nu$ with $-\nu$ in (\ref{eq:BPScigarnu}) 
is clearly a symmetry transformation of this equation
and therefore  indeed in addition to the cigar solution 
there is also a {\it trumpet} solution of the form
\be
e^\nu =  \coth \left( \half \sqrt{c} \tau \right) \qquad  e^{2 \phi} =
\frac{1}{2 a} \frac{1}{ \sinh^2 \left(\half \sqrt{c} \tau \right)}
\end{equation}
Hence the  trumpet solution is the T-dual solution of the cigar
solution.  
Both solutions are drawn in Fig. \ref{Bell}.

\subsection{SUGRA background with a four dimensional  compact transverse space
 ($k = 4$)}

The main difference between the $S_1$ compact transverse space and the general $S_k$ case
is the curvature contribution to the potential in the latter.
Still with no RR or NS form the potential takes the form 
\be
V =  - k(k-1) e^{-2 \nu - 2 \varphi } - c e^{-2 \varphi}
\end{equation}
Let us first check whether the cylinder and cigar solutions can be
generalized to the higher dimensional cases. For that purpose we assume that 
the superpotential  has the form $W(\varphi,\nu)$ so that the relation between the potential and 
the superpotential has the following form
\be
\frac{1}{k} W_\nu^2 - W_\varphi^2 = 16 V.
\end{equation}
Again we take an ansatz  for the superpotential of the form 
$W = 4 e^{-\varphi} w(\nu)$. 
which implies that 
\be
\frac{1}{k} w_\nu^2 - w^2 = -k(k-1)e^{-2 \nu}-c,
\end{equation}
For $k=0,1$ solutions were presented in the previous subsection. 
We were not able to derive a solution for general $k\neq 0,1$ apart from the special case
of a four sphere $k=4$.
It is straightforward to verify that for this case the following is a
solution for the superpotential
\be
w(\nu)= 6 \sqrt{c} e^{-2 \nu} + \sqrt{c}
\end{equation}
As was done above  we  transform to 
 $\tau$ dependence rather than the $\rho$ dependence. 
In terms of this parameterization the BPS equations
are
\be
\pa_\tau \nu  =  \frac{3}{\sqrt{c}}  e^{-2 \nu},
\qquad
\partial_\tau\varphi = -\left( {6}{\sqrt{c}} e^{-2 \nu} + \sqrt{c} \right).
\end{equation}
Finally the solution for  $\nu$ and $\varphi$ is
\be
e^{2 \nu} =   6 c^{-1/2} \tau
\quad
\textrm{and}
\quad
\varphi =  - \ln \tau - \sqrt{c} \tau + A.
\end{equation}
Now since in this solution $\lambda$ is a constant which again we take to be  $\lambda=0$
the metric of this non-critical supergravity  takes the form

\be
ds^2 = dx_{\|}^2 + d \tau ^2 +  6 c^{-1/2} \tau d \Omega_4^2,
\end{equation}
Recall that the range of $\tau$ is  $0 < \tau < \infty$.
The corresponding scalar curvature is given by :

\be
R  = 8 \nu{''} + 20 {\nu{'}}^2 - 12 e^{-2 \nu} = 
                  \frac{1 - 2 \sqrt{c}\tau}{\tau^2},
\end{equation}
which is singular at $\tau \to 0$.

The dilaton on the other hand is regular everywhere as can be seen from the 
following expression 
\be
2 \phi = \varphi + 4 \nu,
\end{equation}
and hence 
\be
e^{2 \phi} \sim \tau e^{-\sqrt{c} \tau}.
\end{equation}


\section{ Conformal   $AdS_{n+1}\times S^k$ backgrounds }
 \label{Conformal}

Next we consider non-critical backgrounds that incorporate RR forms. 
Maldacena's duality  in its original form relates the $AdS_5\times
S^5$ critical string (supergravity  
background)  with the of ${\cal N}=4$ SYM conformal field theory that lives on the
boundary of the  $AdS_5$ (in a large N limit)
.  The $AdS_5\times S^5$ solution was derived as a near horizon 
limit of the supergravity background of $N$ D3 branes, 
but obviously one can generate this solution directly with no reference to the near horizon limit. 
In the search for non-critical  supergravity backgrounds that admit gauge theory duals, 
we will follow the latter approach. 
Following the steps of the  holographic duality for critical strings, we first look for
 ``conformal non-critical backgrounds'' namely non-critical backgrounds 
 with a constant dilaton. 

A brief  glance over (\ref{EMrhophi}) tells us that  requiring  a
constant dilaton implies also a constant $\nu$
\be
\pa_\rho\phi = 0  \qquad \rt \qquad  \pa_\rho\nu = 0
\end{equation}
and the solution of this condition takes the form

\begin{eqnarray}       \label{eq:phi0nu0}
e^{2 \phi_0} &=& \frac{1}{n+1-k} \left( \frac{(n+1-k)(k-1)}{c}
\right)^k\frac{2c}{Q^2}   
 \nonumber \\
e^{2 \nu_0}  &=& \frac{(n+1-k)(k-1)}{c}.    
\end{eqnarray}
In order not to have infinite string coupling or vanishing warp factor
 of the world-volume coordinates,  we must require
\be
n+1-k \neq 0 \qquad k \neq 1.
\end{equation}
It is convenient at this stage to 
switch  from $\rho$ to $\tau$ dependence.  Recalling that $\varphi=2 \phi - n\lambda - k \nu$ 
we find that  the equation for $\lambda$ is 

\be
\pa_\tau^2{\lambda} + n (\pa_\tau{\lambda})^2 = \frac{Q^2}{2} e^{2 \phi_0 - 2k \nu_0}.
\end{equation}  
This is solved by

\be
\lambda = \left( \frac{c}{n(n+1-k)} \right)^{1/2} \tau +\lambda_0.
\end{equation}
It is easy to check that this solution  is in accordance with the zero energy condition.

Defining $R_{AdS}^{1/2} u = e^{\tau R_{AdS}^{-1/2}}$ we end up with the
following metric:

\be
l_2^{-2} ds^2 = ds^2_{AdS_{n+1}} + ds^2_{S^k} 
 =  \left( \frac{u}{R_{AdS}} \right)^2  dx_{\|}^2 
        + \left(\frac{R_{AdS}}{u} \right)^2 du^2 + R_{S^k}^2 d \Omega_k^2,
\end{equation}
where

\be           \label{eq:AdSSRadii}
R_{AdS} = \left( \frac{n(n+1-k)}{c} \right)^{1/2}
\quad \textrm{and} \quad
R_{S^k} = \left( \frac{(n+1-k)(k-1)}{c} \right)^{1/2}.
\end{equation}
The implications of this result is that for any $d$ dimensional space-time there are several
solutions of the form $AdS_{n+1}\times S^k$ such that $n +1 \neq k$
and $k \neq1$.
For instance for $d=6$ the  space-time may be of the following types: 
\be
AdS_{6},  \quad AdS_{4}\times S^2, \quad AdS_{2}\times S^4
\end{equation}
or from a different point of view, four dimensional boundary space-time associated with $AdS_5$
can be accompanied by $S^3$ in addition to the well known 
solution with  $S^5$ if we restrict ourself to even $d$ (as will be
required in Section \ref{holography}) or also $S^0,S^2$ and $S^4$ for general $d$.

Anticipating the discussion of the dual gauge theory (see Section \ref{holography}) it is remarkable that
$e^\phi N$, which will map into  't Hooft coupling constant of the dual gauge theory, appears to
be $Q$-independent:

\be      \label{eq:tHooft}
  g_s N = e^\phi N = 
   \left(  \frac{2 c}{n+1-k} \left( \frac{(n+1-k)(k-1)}{c} \right)^k \right)^{1/2},
\end{equation}
where we have ignored the numerical factors in the relation between  $Q$ and $N$.

\subsection{ The  $AdS_n\times S^k$ backgrounds from BPS equations}

Let us now prove that the solution presented above is supersymmetric, namely,
that it can be  derived  from a  superpotential. (Strictly speaking,  
being a solution of the BPS equation 
is not a sufficient condition for a supersymmetric
 solution but it clearly is a necessary condition). 
If there is a superpotential that  generates the potential, then the
profiles of $\phi$ and $\nu$ are determined by the BPS equations. Since we require 
a constant dilaton, which as we saw implies a constant $\nu$,
we  have to show that at $\phi=\phi_0$ and $\nu=\nu_0$ 
\be
\pa_\phi W|_{\phi=\phi_0, \nu=\nu_0}=0  \qquad \pa_\nu W|_{\phi=\phi_0, \nu=\nu_0}=0 
\end{equation}
If we expand the superpotential around $\phi_0$ and $\nu_0$ this means that there should not be
any liner terms in $\phi$ and $\nu$.

It is convenient to introduce new variables $x$ and $y$:

\be
x = \frac{1}{n-1}(\lambda + \varphi) \qquad \textrm{and} \qquad
y = \frac{1}{n-1}(n \lambda + \varphi) = \frac{1}{n-1}(2 \phi - k \nu).
\end{equation}
Note that $y$ is a constant for the case of our interest since both
$\phi$ and $\nu$ are.
In terms of $\nu$, $x$ and $y$ the superpotential equation reads:

\be
-\frac{1}{n(n-1)} W_x^2 + \frac{1}{n-1} W_y^2 +  \frac{1}{k} W_\nu^2 =
 16 e^{-2nx} \left( Q^2 e^{(n+1)y-k\nu} -c e^{2y} - k(k-1)e^{2(y-\nu)} \right)    .
\end{equation} 
One may simplify this equation substituting $W = -4 e^{-nx} w(y,\nu)$ :

\be   \label{eq:f2}
-\frac{n}{n-1} w^2 + \frac{1}{n-1} w_y^2 +  \frac{1}{k} w_\nu^2 =
  Q^2 e^{(n+1)y-k\nu} -c e^{2y} - k(k-1)e^{2(y-\nu)}  .
\end{equation}
It is remarkable that expanding the expression on the r.h.s. of 
(\ref{eq:f2}) around the point $y=y_0=\frac{1}{n-1}\left(2 \phi_0 -k \nu_0\right)$ and 
$\nu=\nu_0$, where $\phi_0$ and $\nu_0$ are given in (\ref{eq:phi0nu0}), we do not obtain
any term linear in $(y-y_0)$ or $(\nu-\nu_0)$:

\begin{eqnarray}
&& Q^2 e^{(n+1)y_0 -k\nu_0} \Big( -\half (n-1) + \half (n^2-1) (y-y_0)^2 
 \\
&& \qquad  -k(n+3) (y-y_0) (\nu-\nu_0) + \half(k-2) (\nu-\nu_0)^2  \Big)
  + \textrm{higher order terms}.    \nonumber
\end{eqnarray}
It immediately follows that the function $w(y,\nu)$ has a similar form, namely:

\bea
w(y,\nu)  &=& \frac{n-1}{(2n)^{1/2}} Q e^{\half ((n+1)y_0-k\nu_0)} + 
\nonumber \\
&&
 + \left( \half A (y-y_0)^2 + B (y-y_0) (\nu-\nu_0) + \half C (\nu-\nu_0)^2  \right) + \ldots
\eea
for some constant $A$, $B$ and $C$. Plugging this result into the equation of motion 
of $\lambda$, $\nu$ and $\varphi$ we find the solution given in the beginning of this section.
Explicitly, $\dot{\nu}=\dot{\phi}=0$ since $W_\nu$ and $W_\phi$ vanish at 
$\nu=\nu_0$ and $\phi=\phi_0$, while for $\lambda$ we get $\dot{\lambda}= R_{AdS}^{-1}$
in agreement with the previous result.

\subsection{AdS black hole solutions}
\label{AdSBH}

In is well known that on top of extremal  SUGRA backgrounds (in
critical dimensions), one can construct near extremal solutions which
correspond to boundary field theories at finite temperature. 
In particular such backgrounds were written down  for  the near extremal $Dp$ branes
~\cite{Itzhaki:1998dd}. For $D3$ brane in the near horizon limit  the near extremal solution is the 
AdS black hole solution. Since we have identified a family of  $AdS_{n+1}\times S^k$ backgrounds
we would like to examine now whether one can turn them into near
extremal solutions, namely, determine
non-critical AdS black hole solutions.

For this purpose we consider now the following generalization of our initial ansatz:

\be
l_s^{-2} ds^2 = - e^{2 \tilde{\lambda}(\tau)} dt^2 + \sum_{i=1}^{n-1} e^{2\lambda(\tau)} dx_i^2 
       + d\tau^2 +e^{2\nu(\tau)} d\Omega_k^2
\end{equation}
Obviously such solutions will be  non-supersymmetric and hence there is no sense to 
explore them via the BPS equations but rather by 
 the 2nd order equations of motion. We look for solutions 
with constant $\nu$ and $\phi$. 
We first solve  the equations for  $\nu$ and $\phi$
exactly as in the supersymmetric $\lambda=\tilde{\lambda}$ case 
and then we arrive at the following equations
for $\lambda$ and $\tilde{\lambda}$:

\be
\partial_\rho^2 \tilde{\lambda} = \half Q^2 e^{2(\tilde{\lambda}+ (n-1) \lambda - \phi_0)}
\qquad \textrm{and} \qquad 
\partial_\rho^2 \lambda = \half Q^2 e^{2(\tilde{\lambda}+ (n-1) \lambda - \phi_0)}
\end{equation}
The most general solution of this system is:

\be
\tilde{\lambda} = -\frac{1}{n} \ln \left( \frac{1}{a} \sinh(a\alpha\rho) \right) + (n-1) b\rho
\qquad \textrm{and} \qquad
\lambda = -\frac{1}{n} \ln \left( \frac{1}{a} \sinh(a\alpha\rho) \right) - b\rho
\end{equation}
where $\alpha^2 = \half n Q^2 e^{-2 \phi_0}$ and $a$ and $b$ related through 
the zero energy condition:

\be
b =- \frac{\alpha}{n} a.
\end{equation}
In order to rewrite the metric in a more familiar form we will use a new radial coordinate 
$u$ defined by:

\be
e^\lambda = \frac{u}{R_{AdS}}.
\end{equation}
In terms of this coordinate the metric might be easily re-written as:

\be                \label{eq:bh}
l_s^{-2} ds^2 =
   \left( \frac{u}{R_{AdS}} \right)^2 
      \left[ - \left(1 - \left( \frac{u_0}{u} \right)^n \right)  dt^2 +  dx_i^2 \right]  
    + \left(\frac{R_{AdS}}{u} \right)^2
                   \frac{du^2}{ \left(1 - \left( \frac{u_0}{u} \right)^n \right)   }  
    + R_{S^k}^2 d \Omega_k^2,
\end{equation}
where the energy density on the brane is given by $u_0^n= 2a R^n_{ADS}$.
Obviously for zero energy density ($a=0$) we are back in the extremal supersymmetric solution.
The holographic interpretation of these near  extremal solutions will
be addressed in Section \ref{holography}.

\section{The RR deformed  two dimensional black hole}

\label{RRdef2dimBH}

The next goal of our program is to look for non-critical, non
conformal backgrounds, 
namely with non-constant dilaton,  with non-trivial RR forms. 
We start this journey with a two dimensional
non-critical SUGRA background. 

Recall  (\ref{SRR}) that the contribution of the RR charge to the
potential has the form  $Q^2 e^{n \lambda -k\nu- \varphi}$.
In the case of no transverse compact space $k=0$ this is just the potential 
 of a  cosmological
constant $\Lambda$ with $\Lambda = - Q^2 > 0$.
Altogether the potential now reads

\be
V = Q^2 e^{n \lambda - \varphi} - c e^{-2 \varphi}.
\end{equation}

The superpotential is defined via  equation  (\ref{potsup}) which now is 
 $\frac{1}{n} W_\lambda^2 - W_\varphi^2 = 16 V$.
As we have done previously we now factor out $e^{-\varphi}$ so that 
 $W=4 e^{-\varphi} w(\phi)$. We will see later that in general
it is convenient to take $W=4 e^{-\varphi} w(z)$ where 
$z= n\lambda-k\nu +\varphi$ which in our case reduces to $z= n\lambda+ \varphi = 2\phi$.
For $n=1$ the equation for $w(\phi)$ is given by

\be \label{fBGV}
 w^\prime(\phi) w(\phi) - w(\phi)^2 = Q^2 e^{2\phi} - c.
\end{equation}
This equation has an analytic solution ~\cite{Berkovits:2001tg}:

\be
w(\phi) = \sqrt{ 2\phi Q^2 e^{2\phi} -4 m e^{2\phi} + c },
\end{equation}
where $m$ is an integration  constant. From the BPS equations we have:

\be
\frac{\pa \lambda}{\pa \tau} = \half w^\prime(\phi)
\qquad
\frac{\pa \varphi}{\pa \tau} = w(\phi) - \half w^\prime(\phi),
\end{equation}
which implies in particular that:

\be
 \frac{\pa \phi}{\pa \tau} = \half  w(\phi).
\end{equation}
It turns out that it is useful to take 
$\phi$ as a radial coordinate instead of $\tau$.
It is easy to check that in this parameterization 
$e^{2\lambda}= w^2$ so that we 
we end up with the 2d black hole metric of ~\cite{Berkovits:2001tg} 

\be        \label{eq:BGVphi}
l_s^{-2} ds^2 = -\frac{1}{4} w^2(\phi) dt^2 + \frac{d \phi^2}{\frac{1}{4}w^2(\phi)},
\end{equation}
In ~\cite{Berkovits:2001tg} the metric is expressed in terms of  $l(\phi)=\frac{1}{4} w^2 ( \phi)$.
It was shown in ~\cite{Berkovits:2001tg} that this solution can be interpreted as 
a two dimensional black hole  with an ADM mass
$M_{ADM}=\frac{2}{\sqrt{c}}m$. The scalar curvature was found to be
$\alpha^\prime \mathcal{R}=  e^{2 \phi}[Q^2(\phi+1)-m]$.

In fact using $\phi$ as the radial coordinate, one can derive this solution
directly from the equations of motions.
In Appendix \ref{BGVsolution} we find that the function 
$e^{2\lambda}= l(\phi)$ 
is indeed the same as derived from the BPS equations.
The points $\phi=\phi_0$ where $w(\phi)=0$ constitute the  horizon. 
In the extremal case $l(\phi)$ has a double zero at $\phi=\phi_0$
and the space-time geodesics are complete (the integral $\int\frac{d \phi}{l^{\half}(\phi)}$
diverges at $\phi_0$).
The double zero requirement leads to 

\be \label{eq:phi0a}
e^{2\phi_0} = \frac{c}{Q^2}
\qquad
\textrm{and}
\qquad
 m = \frac{1}{4} Q^2 (1+2 \phi_0).
\end{equation}
The function $l(\phi)$, which satisfies the above conditions is
plotted in Fig. \ref{BGV}.

\FIGURE[b]{
 \label{BGV}
\centerline{\begin{picture}(0,0)%
\includegraphics{BGV.pstex}%
\end{picture}%
\setlength{\unitlength}{2960sp}%
\begingroup\makeatletter\ifx\SetFigFont\undefined%
\gdef\SetFigFont#1#2#3#4#5{%
  \reset@font\fontsize{#1}{#2pt}%
  \fontfamily{#3}\fontseries{#4}\fontshape{#5}%
  \selectfont}%
\fi\endgroup%
\begin{picture}(3600,2631)(751,-2692)
\put(2251,-286){\makebox(0,0)[lb]{\smash{{\SetFigFont{14}{16.8}{\rmdefault}{\mddefault}{\updefault}$l(\phi)$}}}}
\put(4351,-2611){\makebox(0,0)[lb]{\smash{{\SetFigFont{14}{16.8}{\rmdefault}{\mddefault}{\updefault}$\phi$}}}}
\end{picture}%
}
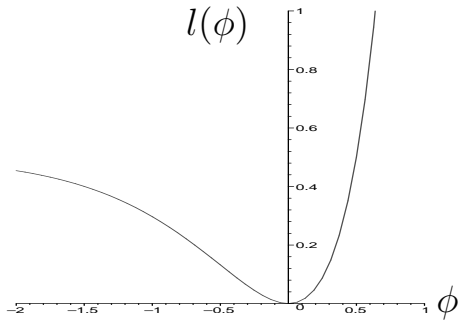
\caption{The function $l=l(\phi)$ for $Q^2=2$, $m=\half$ and
  $\phi=\phi_0=0$. Note that
  $l(\phi)^\prime=l(\phi)=0$ at $\phi_0=0$ in agreement with (\ref{eq:phi0a}).}
}

In ~\cite{Berkovits:2001tg} the near horizon limit of the black hole solution was shown to
be a two dimensional AdS solution. Expanding $l(\phi)$ around $\phi_0$ takes
the form $l(\phi)\sim \frac{c}{2}(\phi-\phi_0)^2$ since 
$l(\phi_0)= l'(\phi_0)=0$. Defining now a new variable
 $u = \phi-\phi_0$ the near horizon metric is 

\be           \label{eq:AdS2}
l_s^{-2} ds^2 = - \left(\frac{u}{R_{AdS}} \right)^2  dt^2  + \left( \frac{R_{AdS}}{u} \right)^2 du^2,
\end{equation}
where $R_{AdS}=\sqrt{\frac{2}{c}}$ and the scalar curvature is given by
$ \alpha^\prime \mathcal{R}_{AdS} = {c}$.
It will be useful for later purposes to find this metric directly from the 
second order equations of motion.

\be
\pa_\rho^2 \lambda = \half Q^2 e^{\lambda -\varphi},
\qquad
\pa_\rho^2 \varphi = - c e^{-2 \varphi} + \half Q^2 e^{ \lambda -\varphi},
\end{equation}
thus

\be
\pa_\rho^2 ( \lambda+ \varphi) = (-c + Q^2 e^{\lambda +\varphi} ) e^{-2 \varphi}.
\end{equation}
This equation is solved by (\ref{eq:phi0a}) in accord with the previous discussion.
Plugging this result into the equation for $\varphi$ we get:

\be   \label{d2varphi1}
\pa_\rho^2 \varphi = - \frac{c}{2} e^{-2 \varphi}. 
\end{equation}
This is easily solved by $e^\varphi =\sqrt{\frac{c}{2}} \rho$ and therefore 
$\varphi=\sqrt{\frac{c}{2}} \tau$. Using the result for the dilaton we find that:

\be
e^{ \lambda} =  \sqrt{\frac{c}{2}} e^{\sqrt{\frac{c}{2}} \tau}.
\end{equation} 
Defining 
$u = e^{\sqrt{\frac{c}{2}} \tau }$
we arrive at the AdS metric.

Few remarks are in order:

\begin{itemize}

\item
 In fact this type of solution with constant dilaton exists also for general $n$.
It is easy to verify that the corresponding solution is
\be
e^{2\phi_0}= \frac{2}{n+1}\frac{c}{Q^2}
\qquad 
\lambda =   \left(c \frac{n}{n+1}\right)^\half   \tau + \lambda_0
\end{equation} 
Defining 
$u = e^{\left(c \frac{n}{n+1}\right)^\half \tau }$
we arrive at the following AdS metric \footnote{For $n=1$ this result appears in BGV}:

\be    \label{eq:AdSn}
l_s^{-2} ds^2 = \left( \frac{u}{R_{AdS}} \right)^2 dx_\mu^2 
            + \left(\frac{ R_{AdS}}{u} \right)^2 du^2,
\end{equation}
where $R_{AdS} =  \sqrt{\frac{n(n+1)}{c}}$. 

\item
Note that since the string coupling is $g_s= e^{\phi_0}= \frac{2}{n+1} \frac{\sqrt{c}}{Q}$,
large RR charge   $Q$  means weak  string coupling  so that  we can trust
the perturbative string description. 

\item 
On the other hand the since the scalar curvature  is 
$R=\frac{c}{n(n+1)}$ it does not depend on the number of branes and remains fixed in the large
$N$ limit. This is obviously  not like for 
the $AdS_5 \times S_5$ case, where the small curvature requires 
large $N$. 
In fact this is a particular AdS solution discussed in  Section \ref{Conformal}.

\end{itemize}


\subsection{ The T-dual solution of the BGV black hole $k=1$}

Consider now instead of a space-time with one world volume coordinate
and no compact transverse space $(k=0, n=1)$ the  two dimensional case
with no world volume coordinate but 
with an $S^1$ direction namely $(k=1, n=0 )$.  Recall that the potential for $k=1$ takes the form

\be
V= Q^2 e^{-  \nu - \varphi}  -c e^{-2 \varphi}
\end{equation}
Using again the  following parameterization of  the  superpotential

\be
W = 4  e^{-\varphi} w(z), \quad \textrm{where} \quad 
  z =  -\nu+\varphi.
\end{equation}
The superpotential equation reads

\be     \label{ODEk1}
2 w^\prime(z) w(z) - w(z)^2 = Q^2 e^{z} - c
\end{equation}
and therefore

\be      \label{eq:BGVlike}
w^2(z) = Q^2 ze^{z} - A e^{z} + c, 
\end{equation}
where $A$ is constant. It means that  the metric is

\be      \label{eq:BGVdual}
l_s^{-2} ds^2 = \frac{1}{w^2(z)} \left( dz^2 + d \theta^2 \right),
\end{equation} 
where $d \tau = \frac{d z}{w(z)}$ and the dilaton is given by $e^{2 \phi} \sim w^{-2}(z)$.
In fact it is easy to realize that this $(k=1, n=0 )$ solution is just the 
T-dual  of the BGV solution.  
As discussed for backgrounds with no RR charges 
T-duality implies inverting the radius of the compact direction and transforming
the dilaton $e^\phi \rt \frac{e^\phi}{g_{\textrm{compact}}}$ 
where $g_{\textrm{compact}}$ is the metric 
along the compact direction. For the cigar/trumpet duality these transformation 
were $\nu\ \rt \  -\nu$ and $e^{2\phi} \rt e^{2\phi-2\nu}$.
In the present case the situation is slightly more complicated due to the RR flux.
The BGV background is equipped with a RR zero form  
with  one dimensional world volume. Taking the latter 
to be compact and performing T duality  along this circle
transforms the RR flux into a 1-form RR flux  with no world 
volume but with additional transverse compact one dimensional
 transverse space. This is indeed the passage 
from $(k=0, n=1)$  to   $(k=1, n=0 )$. Now the original world volume 
compact coordinate with 
metric $e^{2\lambda}$ is transformed into a compact 
transverse direction  with metric   $e^{2\nu}=e^{-2\lambda}$ and 
$g_{\textrm{compact}}$ for the dilaton transformation is
$g_{\textrm{compact}}=e^{2\lambda}$. 
Thus T-duality is summarized in the following transformations
\footnote{Note that the we also modify the signature of the metric.}
\be
(k=0, n=1)\  \rt\  (k=1, n=0 )\qquad \lambda\ \rt \ 
   -\nu \qquad   e^{2\phi} \rt e^{2\phi-2\nu}.
\end{equation}

Similarly to the the BGV solution the function $w(z)$ in
(\ref{eq:BGVlike}) depends on the integration constant $A$.
In (\ref{eq:BGVphi}) this constant was fixed by requiring that the metric
should be geodesically complete near the horizon defined by $w(z_0)=0$.  
After T-duality, however, the horizon is removed to $z=\infty$ and at the point
$w(z_0)=0$ one has instead a diverging string
coupling. Therefore we would like to require that $w(z) \neq 0$
for any $z$. It leads to (see Fig. \ref{BGVdual} for 
$w^{-2}(z)$ vs. $z$ for different values of $A$):

\FIGURE[b]{
 \label{BGVdual}
\centerline{\input{BGVdual.pstex_t}}
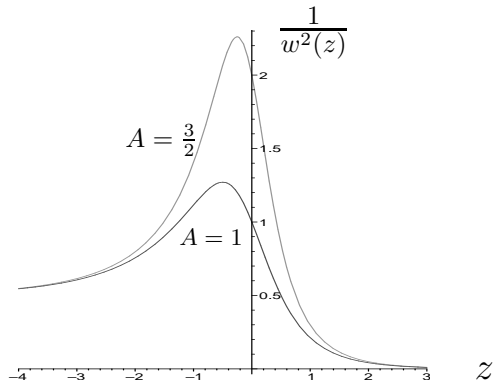
\caption{The function $w^{-2}(z)$ for $Q^2=2$ and  $A=1,\frac{3}{2}$.
}
}

\be
A \leq A_{crit} = Q^2 \left(1+ \ln \frac{c}{Q^2} \right).
\end{equation}
For $A < A_{crit}$ the solution interpolates between the cylinder at
$z \to -\infty$ and a horizon at $z \to \infty$, where $w^{-2}(z) \approx 0$. 
The curvature of the solution is given by 
$\alpha^\prime \mathcal{R} = 2 ( {w^\prime}^2 -w^{\prime\prime} w)$ and
it diverges for large $z$. The string coupling $e^\phi \sim
\frac{1}{w(z)}$, however, is finite
everywhere getting its maximal value at the point $z=z_0$
satisfying $w^\prime(z)=0$.

It is worth to remark that if the above inequality is precisely
satisfied ($A=A_{crit}$),
then near the point $w(z_0)=0$ the metric (\ref{eq:BGVdual}) reads:

\be
l_s^{-2} ds^2 = \left( \frac{R_{AdS}}{u} \right)^2 (du^2 + d \theta^2)
\end{equation}
and the dilaton behaves like $e^{2 \phi} \sim  u^{-2}$.
To derive this result we have applied the T-duality rules on the AdS 
metric (\ref{eq:AdS2}). Remarkably the new metric again describes
$AdS_2$ (with a non-constant dilaton and Euclidean signature).
Unfortunately , at $u \to 0$ the string coupling diverges making the supergravity
description irrelevant.

\section{Backgrounds with non-zero RR charge $Q \neq 0$ that asymptote to
  the linear dilaton solution} 
\label{RRdef}

In section Section \ref{k=1} the non-critical cigar and trumpet  backgrounds were determined 
for  cases with  no RR flux. Let us now examine the  fate of these
backgrounds once we turn on a RR flux. As we will see in Section \ref{holography}
these SUGRA backgrounds will turn out to be 
 of special interest in the search of duals of the ${\cal N}=1$ SYM gauge theories.
To  study the deformation of the cigar solution,
 we have to investigate the superpotential equation (\ref{potsup})
with $k=1$, $Q\neq 0$ and arbitrary $n$.
Recall our parameterization of the superpotential 
$W = 4  e^{-\varphi} w(z)$ where   
 $ z =n \lambda-\nu+\varphi$ such that

\be     \label{eq:ODEk11}
n {w^\prime(z)}^2 + 2 w^\prime(z) w(z) - w^2(z) = Q^2 e^{z} - c
\end{equation}
In this parameterization the BPS equations read

\begin{eqnarray}  \label{eq:lnfz}
 \frac{\partial \lambda}{\partial \tau} =  w^\prime(z),
\qquad
\frac{\partial \nu}{\partial \tau} = - w^\prime(z),
\qquad 
 \frac{\partial \varphi}{\partial \tau} = - w^\prime(z) + w(z).
\end{eqnarray}
Using these identities we can write the equation for $z$:

\be        \label{eq:dzdtau}
\frac{\partial z}{\partial \tau} =  n w^\prime(z) + w(z).
\end{equation}
We will now use   $z$ as the radial coordinate so that 
the general form of the metric
is:

\be   \label{eq:metricfz}
l_s^{-2} ds^2 = e^{2 \lambda(z)} dx_\mu^2 + \frac{dz^2}{ \left( n w^\prime(z) + w(z) \right)^2} +
          e^{2 \nu(z)} d\theta^2.
\end{equation}

The non-linear equation  (\ref{eq:ODEk11}) has analytic solutions only
for $n=0$ (see the discussion following (\ref{ODEk1})) and $n=-1$, with the latter being unphysical. 
Since we do not have an analytic solution let us first discuss the 
asymptotic behavior of the background. For the latter to correspond to a cylinder
metric and a linear dilaton, the superpotential should asymptote as follows
$W\rt \pm \sqrt{c} e^{-\varphi}$, namely, $w(z)\rt \pm \sqrt{c}$. 
Taking $z\rt -\infty$ as the asymptotic region, (\ref{eq:ODEk11}) implies that asymptotically 
$w(z)'\rt 0$. 
We will choose the ``$-$'' solution for a later convenience. 
\footnote{The equation (\ref{ODEk1}) is a second degree differential equation and generally
there are two families of solution corresponding to the two roots of the second degree equation:

\be    \label{eq:wzprime}
w^\prime = -\frac{1}{n} \left( w \pm \sqrt{(n+1)w^2 + n (Q^2 e^{z} - c) } \right).
\end{equation}
The boundary conditions $w(z)\vert_{z \to \infty} \to \sqrt{c}$  and
$w^\prime(z)\vert_{z \to \infty} \to 0$ are related to the 
''-'' choice in (\ref{eq:wzprime}).}

The perturbative expansion of $w(z)$ around  $z \to -\infty$ is given by:

\be
w(z) \approx  \sqrt{c} - \frac{1}{2} \frac{Q}{\sqrt{c}}z e^{z}  
\quad \textrm{for} \quad
z \to -\infty
\end{equation}
and we can calculate an approximate solution for the metric and the
dilaton:

\be
\lambda(\tau) \approx -\half \frac{Q}{\sqrt{c}} \tau e^{ - \sqrt{c} \tau}, \quad
\nu(\tau) \approx \half \frac{Q}{\sqrt{c}} \tau e^{ - \sqrt{c} \tau}, \quad
2 \phi \approx \sqrt{c} \tau.
\end{equation}
As expected the warp functions become  constant and the dilaton depends
linearly on $\tau$. These expressions also show that the region
$z \rt -\infty$
corresponds to $\tau \rt -\infty$. 
The function $w(z)$ can be extended numerically to the region with finite $z$.
As it shown on Fig.\ref{wzn4a} there are three possible scenario at
finite $z$:

\FIGURE[b]{
 \label{wzn4a}
\centerline{\begin{picture}(0,0)%
\includegraphics{wzn4a.pstex}%
\end{picture}%
\setlength{\unitlength}{3947sp}%
\begingroup\makeatletter\ifx\SetFigFont\undefined%
\gdef\SetFigFont#1#2#3#4#5{%
  \reset@font\fontsize{#1}{#2pt}%
  \fontfamily{#3}\fontseries{#4}\fontshape{#5}%
  \selectfont}%
\fi\endgroup%
\begin{picture}(5525,4335)(2692,-5611)
\put(7941,-5553){\makebox(0,0)[lb]{\smash{{\SetFigFont{12}{14.4}{\familydefault}{\mddefault}{\updefault}$z$}}}}
\put(5703,-1440){\makebox(0,0)[lb]{\smash{{\SetFigFont{12}{14.4}{\familydefault}{\mddefault}{\updefault}$w(z)$}}}}
\end{picture}%
}
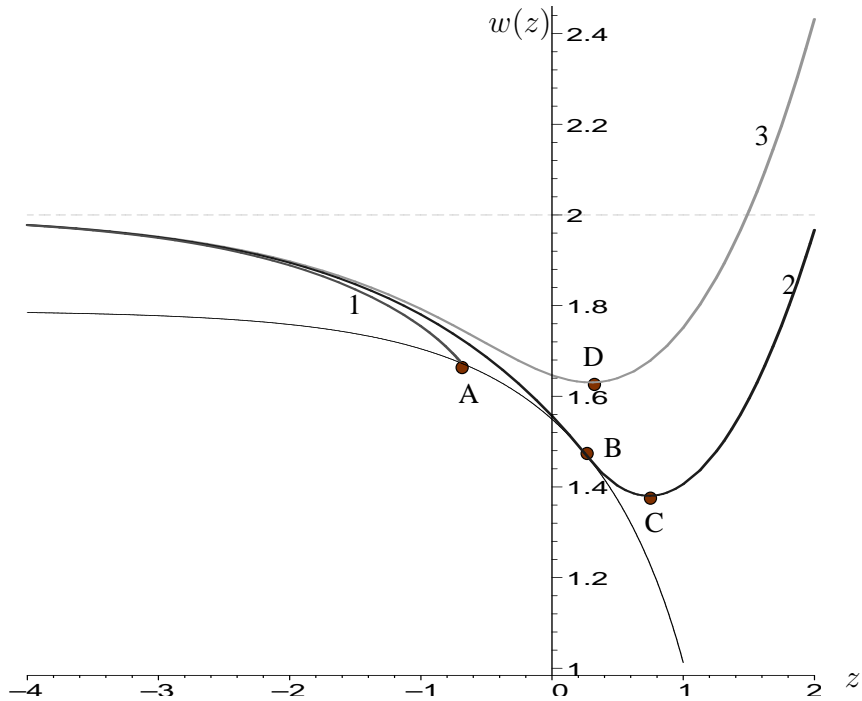
\caption{ The plot represents three possible solutions (solid curves) of the equation
 (\ref{eq:ODEk11}) for $n=4$, $Q=1$. All the solutions approach
 the asymptotic value  $\sqrt c$ (the dashed line) at $z
 \to - \infty$. The numerical expansion related to the curve no.1
 terminates at point A, where it meets the thin curve 
 defined by the (\ref{eq:nww0}) or alternatively by $n w^\prime +w=0$.  
 The solution no.2, however, is only tangent to the $n w^\prime + w =0$ curve at
 point B and therefore can be extended to any value of $z$. Finally,
 the solution no.3 does not meet the thin curve, which means that
 along this curve one always has $n w^\prime + w >0 $. Both no.2 and
 no.3 have $w^\prime =0$ exactly at one point (C and D respectively).
} 
}

\begin{itemize}
\item[1.] This expansion terminates (point A on Fig. \ref{wzn4a}), 
          when we approach the curve described by
          \be  \label{eq:nww0}
           w(z) = \left( \frac{n}{n+1} (c- Q^2 e^{z})\right)^{1/2},
         \end{equation}
         which can be also defined by $n w^\prime(z) + w(z)=0$. 
         Indeed, this is a simple exercise to verify that
          (\ref{eq:ODEk11}) has no solution for $w^2(z) <   \frac{n}{n+1} (c- Q^2 e^{z})$.
\item[2.] The curve that describes the numerical solution (no.2 on Fig.\ref{wzn4a})
          is only tangent at $z=z_0$ (point B on Fig.\ref{wzn4a}) to
          the curve defined by (\ref{eq:nww0}).  In this case the numerical solution
          might be extended to any value of $z$. In particular, at $z \to \infty$
          one has $w(z) \approx \frac{2}{\sqrt{n}} Q e^{z/2}$. 
\item[3.] In this case the curve (no. 3 on Fig. \ref{wzn4a})  does not approach the
          curve  (\ref{eq:nww0}) and the numerical expansion exists for
          any $z$. It is important to no note, that $n w^\prime(z) + w(z) >0$  
          along this solution.
\end{itemize}

Note that from  $n w^\prime(z) + w(z) >0$ follows that $z(\tau)$ is a
monotonic function of $\tau$.
The completeness of space-time geodesics at infinity 
(namely at $z=z_0$ where $n w^\prime(z) + w(z) =0$)
implies that the function $nw^\prime(z)+w(z)$ has a simple zero at $z=z_0$.
After some algebra we find from this condition that:

\be  \label{eq:z0}
e^{z_0} = \frac{2}{n+2} \frac{c}{Q^2},
\quad \textrm{so that} \quad 
w(z_0) = -n w^\prime(z_0) =  \frac{n}{n+1} \alpha, 
\quad \textrm{where} \quad
\alpha^2 =  \frac{n+1}{n+2} c.
\end{equation}
The solution no.1 listed above terminates at $z_A<z_0$ and therefore
the space described by this curve is geodesically
incomplete. Furthermore, this is a straightforward exercise to verify
that at $z=z_0$ the solution of (\ref{eq:ODEk11}) is tangent to the
curve (\ref{eq:nww0}) and as a consequence the function $nw^\prime(z)+w(z)$ has a simple
zero at this point. This is the curve no.2 on Fig.\ref{wzn4a}  and
point B is located precisely at 
$z=z_0$.
Let us describe this solution in more details.
The curvature of the metric (\ref{eq:metricfz}) can be written in terms of the function $w(z)$:

\FIGURE[b]{
 \label{AdSlikeLN}
\centerline{\input{AdSlike.pstex_t}}
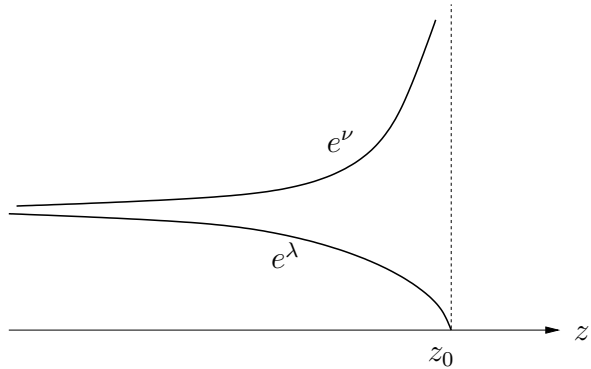
\caption{The factors $e^\lambda$ and $e^\nu$ vs. the coordinate $z$
         for the solution corresponding to the curve no.2 on
         Fig.\ref{wzn4a}. The point $z=z_0$ is mapped to $\tau = - \infty$.
} 
}

\be
\alpha^\prime \mathcal{R} = 2 (n-1) w^{\prime\prime} (nw^\prime +w) + (n^2-n+2) {w^\prime}^2, 
\end{equation} 
where we made use of (\ref{eq:lnfz}). At $z \to -\infty$ (and $\tau \to -\infty$)
the solution reduces to the cigar geometry ($\lambda=\nu=0$) and the curvature goes to zero as
expected. At  $z \to z_0$ (alternatively $\tau \to \infty$) the curvature
has a non-zero value:

\be \label{Rz0}
\alpha^\prime \mathcal{R} =  \frac{n^2-n+2}{(n+1)(n+2)} c.
\end{equation}
In this region we can find the solution using (\ref{eq:z0}):

\be
\lambda = \frac{\alpha}{n+1} \tau, \qquad
\nu = - \frac{\alpha}{n+1} \tau, \qquad
 \varphi =- \alpha \tau + \frac{z_0}{2}
\quad \textrm{and} \quad \phi = - \frac{\alpha}{n+1} \tau + \frac{z_0}{2}
\end{equation}
and therefore there is a horizon at $\tau \to -\infty$.
Defining

\FIGURE[b]{
 \label{e2L}
\centerline{\input{e2L.pstex_t}}
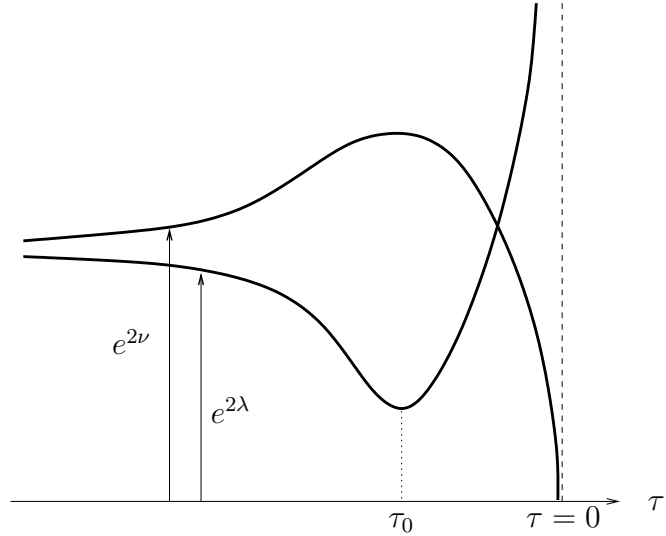
\caption{ The picture represents the typical form of
  $g_{ii}=e^{2\lambda}$ and $g_{\theta\theta}=e^{2\nu}$. For $\tau \to -\infty$
  we approach the cylinder geometry, while at $\tau \to 0$ the
  background becomes singular. 
  The function $e^{2\lambda}$ has a global minimum at $\tau=\tau_0$. 
} 
}

\be
u = e^{\frac{\alpha}{n+1} \tau}
\end{equation}
we end up with the following background:

\be   \label{AdSlike}
l_s^{-2}ds^2 = \left( \frac{u}{R_{AdS}}\right)^2  dx_{\|}^2 
          + \left(\frac{R_{AdS}}{u}\right)^2 du^2
          + \left(\frac{R_{AdS}}{u} \right)^2 d \theta^2,
\quad \textrm{where} \quad
 R_{AdS} =\frac{n+1}{\alpha}
\end{equation}
and the dilaton is:

\be         \label{eq:DivDilaton}
e^{\phi} = \left( \frac{2}{n+2} \frac{c}{Q^2} \right)^{1/2} \frac{1}{u}.
\end{equation}
Actually,  the background in the near horizon region is 
the T-dual version of the AdS solution (\ref{eq:AdSn})  for arbitrary $n$ and $k=0$.
As an additional non-trivial check one can calculate the curvature of the near horizon metric, 
which should coincide with the previously found result (\ref{Rz0}).
To summarize, the background related to the curve no.2 on Fig.\ref{wzn4a} interpolates between
the cylinder solution with a linear dilaton at $\tau \to -\infty$ ($z \to -\infty$)
and a configuration consisting of the metric (\ref{AdSlike}) and the
dilaton (\ref{eq:DivDilaton}) at $\tau \to -\infty$ ($z \to z_0$). 
Unfortunately, the dilaton diverges at $u \to 0$ making
the supergravity description irrelevant. We will return to this
point latter discussing the holographic properties of this solution.   
In contrast with the cigar
solution the warp function of the angular part of the metric diverges
at  $\tau \to -\infty$, while the world volume warp factor goes to
zero (see Fig.\ref{AdSlikeLN}). 
Notice also that the boundary of the metric (\ref{AdSlike}) at $u \to \infty$
is four dimensional exactly like in the $AdS_5 \times S_5$ case.

Let us comment briefly on how one arrives at the near horizon solution directly from the second order
equations of motion. From (\ref{EMrho}) we get for $k=1$:

\be
\pa^2_\rho (n \lambda - \nu + \varphi) = 
 \left( -c+ \half (n+2) Q^2 e^{n \lambda - \nu + \varphi} \right) e^{-2 \varphi}.
\end{equation}
This is trivially solved provided that $z=n \lambda - \nu + \varphi$ is constant
satisfying (\ref{eq:z0}). Furthermore, from the the equation for $\varphi$ we obtain:

\be      \label{xxx}
\pa^2_\rho \varphi = - \alpha^2 e^{-2 \varphi},
\end{equation}
which is solved by $e^\varphi = \alpha \rho$. Plugging this into the equations for 
$\lambda$ and $\nu$ one arrives at the same near horizon background we have found 
above.

It turns out that  there is an additional solution of (\ref{xxx}), which does not appear as 
a near horizon limit of the solution with non-constant $z$:

\be
e^\varphi = \frac{1}{a} \sinh(\alpha a \rho).
\end{equation}
Following the same steps as in the $k=0$ we find

\bea
e^\lambda &=& \left( 4 \sinh^2 \left(\half \alpha \tau \right) \right)^{\frac{1}{n+1}} \\
\nonumber
e^\nu &=& \left( 2 \sinh \left( \alpha \tau \right) \right)^{-\frac{1}{n+1}}
       \left( \tanh \left(\half \alpha \tau \right) \right)^{\frac{n}{n+1}} \\
\nonumber
e^\phi &=& \left(\frac{2}{n+2} \frac{c}{Q^2} \right)^{1/2}
               \left( 2 \sinh \left( \alpha \tau \right) \right)^{-\frac{1}{n+1}}
               \left( \tanh \left(\half \alpha \tau \right) \right)^{\frac{n}{n+1}}. 
\eea
For large $\tau$ the background degenerates to the near-horizon solution we have identified
above.
At $\tau =0$ for $n>1$ the curvature diverges and there is a naked singularity at this point.

Next we will investigate the solution related to the curve no.3.
In this case the function $n w^\prime + w$ does not vanish at any $z$.  
Similarly to the curve no.2 at $\tau,z \to -\infty$ we have $w \approx\sqrt{c}$
and the solution in this region is the cylinder. On the contrary, at  
$z \to \infty$ the solution of (\ref{eq:ODEk11}) is
$w \approx \frac{2}{\sqrt{n}}Qe^{z/2}$, so that $\tau \to 0$ and

\be
e^\lambda \approx \tau^{-\frac{1}{n+2}},
\qquad
e^\nu \approx \tau^{\frac{1}{n+2}}
\qquad \textrm{and}
\quad 
e^\phi \approx \tau^{-\frac{n}{n+2}}.
\end{equation}
The curvature in this region diverges and the string coupling becomes infinite.
At this stage it is important to emphasize that there a is unique point
along the curve no.3 with $w^\prime(z)=0$ (point D on
Fig.\ref{wzn4a}). It immediately follows from (\ref{eq:lnfz}), that
$\lambda(\tau)$ (and also $e^{\lambda(\tau)}$) has a minimum at this
point ($\dot{\lambda}=0$). This new feature will be discussed in
Section \ref{holography} in the context of confiment in the
dual gauge theory. The typical form of the functions $e^\lambda$
and $e^\nu$ is plotted on Fig.\ref{e2L}.

Our original goal in this section was to construct the RR
perturbation of the cigar geometry. As it evident, however, from all the solutions
the asymptotic form of the metric reproduces the cylinder, rather then
the cigar background. In particular, we have not obtained a $e^\nu$ factor that  
monotonically decreases to zero, like in the cigar solution.
It seems that in order to reach the goal,
a more general ansatz for the solution of the superpotential equation
is needed.
For example, one may consider a superpotential of the form
$W=4e^{-\phi} w(z,\nu)$, which seems to be the most simple incorporation
of the cigar solution ($w$ depends only on $\nu$)
and the solution investigated in this section ($w$ depends only on $z$).

Before closing this section let us comment on the T-dual of the
background. According to the discussion in the previous section
under T-duality along the circle 
defined by the coordinate $\theta$  the type IIB metric (\ref{eq:metricfz})
transforms into type IIA metric:

\be     \label{eq:BGVn}
l_s^{-2} ds^2 = e^{2 \lambda(\tilde{z})} dx_\|^2 
     + \frac{d\tilde{z}^2}{ \left( n \tilde{w}^\prime(\tilde{z}) + \tilde{w}(\tilde{z}) \right)^2},
\end{equation}
where $\tilde{z}=(n+1)\lambda+\varphi=2\phi$ and $\tilde{w}$ satisfies an
equation similar to (\ref{eq:ODEk11}):

\be     
(n-1) {\tilde{w}^\prime(\tilde{z})}^2 + 2 \tilde{w}^\prime(\tilde{z})\tilde{w}(\tilde{z}) 
     - \tilde{w}(\tilde{z})^2 = Q^2 e^{\tilde{z}} - c.
\end{equation}
It is important to note that the type of the superpotential equation
solution we have considered in the beginning of the section requires
automatically that $\lambda=-\nu$. So performing the T-duality
transformation we arrive at the metric compatible with our initial
ansatz (\ref{stringmetric}).

Again the equation for $\tilde{w}(\tilde{z})$ has no explicit solution for $n>1$. For $n=1$ we
return to the BGV black hole solution discussed earlier.
For $n>1$ the numerical solutions appear on Fig.\ref{wzn4a},
with the coordinate $z$ replaced by $\tilde{z}=2 \phi$.
Finally, the typical form of the function $e^{2 \lambda(\tilde{z})}$
repeats the result for  $e^{2 \lambda(z)}$
on Fig.\ref{AdSlikeLN} and Fig.\ref{e2L}. 
In the former case in the near horizon region ($e^{\lambda(\tilde{z})} \to 0$)
the solution reduces to the $AdS_n$ background (\ref{eq:AdSn}).



\section{Solutions with a non-trivial NS-NS charge }   \label{QNS}

Next we would like to consider non-critical backgrounds that include 
$H$ the  NS-NS three form but no RR forms. 
In (\ref{SrrSns}) we have assumed that $H$ lives in the compact transverse space. 
That obviously implies that the dimension of the compact transverse space has to be $k=3$.
The potential associated with this case is 

\be
V = Q^2 e^{-2k \nu - 2 \varphi} - 6 e^{-2\nu -2 \varphi} -c  e^{-2 \varphi}.
\end{equation}
Note that unlike the RR case here the potential is independent of $\lambda$. We therefore
 look for a superpotential in the form $W=4 e^{-\varphi} w(\nu)$. We arrive at the following
differential equation for $w(\nu)$:

\be \label{fnu}
\frac{1}{3} {w^\prime}^2 (\nu) - {w}^2 (\nu) = Q^2 e^{-6 \nu} - 6  e^{-2 \nu} -c
\end{equation}
and the BPS equation for $\nu$ and $\varphi$ are

\be \label{BPSk3}
\frac{\partial \nu}{\partial \tau} =\frac{1}{3} f^\prime(\nu),
\qquad 
 \frac{\partial \varphi}{\partial \tau} = w(\nu). 
\end{equation}
Using $\nu$ as the radial coordinate instead of $\tau$ we may rewrite the metric as:

\be
ds^2 = dx_{\|}^2 + \frac{9}{{w^\prime}^2(\nu)}d \nu^2 + e^{2 \nu} d \Omega_3^2.
\end{equation}
It is clear therefore that the at $\nu=\nu_0$ satisfying $w^\prime(\nu)=0$ 
there is a horizon manifold. 

As another warmup exercise let us  analyze first  the critical case ($c=0$). 
In this case the equation (\ref{fnu}) has two distinct solutions:

\be
w(\nu) = \frac{1}{\sqrt{2}} Q e^{-3 \nu} + 3 e^{-\nu}
\qquad \textrm{and} \qquad 
f(\nu) = \frac{1}{\sqrt{2}} Q e^{-3 \nu} - 3 e^{-\nu}.
\end{equation}
In the first case $w^\prime(\nu)$ do not vanish anywhere and the
corresponding metric 
describes a space with a boundary at $\nu \to -\infty$.
In the later case $w^\prime(\nu)=0$ exactly at one point $\nu=\nu_0$ and substituting 
this function in the BPS equations (\ref{BPSk3}) for the case of $k=3$
we will obtain the well known NS5 solution.
In particular, near $\nu=\nu_0$ it is described by the linear dilaton configuration.
Unfortunately we cannot find an analytic solution of (\ref{fnu}) in the non-critical case.
We will fix the boundary condition at $\nu \to \infty$ by $w^\prime(\nu) \to 0$.
Then the asymptotic behavior of $w(\nu)$ at $\nu \to \infty$:

\be  \label{fnuapprox}
w(\nu) \approx -\sqrt{c}- \frac{3}{\sqrt{c}} e^{-2 \nu} + \ldots
\end{equation}
so that at large $\tau$ we get:

\be
e^{2 \nu} \approx 4 c^{-\half} \tau
\quad \textrm{and} \quad 
2 \phi \approx \varphi \approx - c^{-\half} \tau.
\end{equation}
The solution (\ref{fnuapprox}) can be interpolated numerically to the finite $\nu$ region.
At some point $\nu=\nu_0$ this extension will approach
the curve 

\be
f(\nu)=-\sqrt{Q^2 e^{-6 \nu} - 6 e^{-2 \nu} -c},
\end{equation}
where $f^\prime(\nu)=0$. We can find  $\nu_0$ by demanding that  $f^\prime(\nu)$
has a single zero at this point and therefore the space-time geodesics
are complete. It results in:

\be
e^{4 \nu_0} = \frac{Q^2}{2}
\quad \textrm{and} \quad
f(\nu_0) = -\sqrt{c + 4 \frac{2^{\half}}{Q}} \equiv - \gamma.
\end{equation}
Finally, at  $\tau \to -\infty$  we have:

\be    \label{NS5like}
\nu = \nu_0
\quad \textrm{and} \quad
e^{2 \phi} \sim e^{- \gamma \tau}, 
\end{equation}
which is the standard linear dilaton solution one obtains in 
the decoupling limit of the NS5 background.
The curvature is given in terms of $f(\nu)$ by:

\be
\alpha^\prime \mathcal{R} = 2 w^{\prime\prime} w^{\prime} -\frac{4}{3} {w^\prime}^2 - 6e^{-2 \nu_0} 
\end{equation}
As expected it goes to zero at large $\nu$ ($\tau \to \infty$) and at
for $\nu \to \nu_0$ 
($\tau \to -\infty$) one finds $\alpha^\prime \mathcal{R} \approx \frac{6 \sqrt{2}}{Q}$. 

Again one can derive the near horizon solution directly from the 
second order differential equations:

\be
\pa^2_\rho \lambda = 0 , \quad
\pa^2_\rho \nu - 2 e^{-2\nu-2\varphi} + Q^2 e^{-6\nu-2\varphi} = 0 
\end{equation}
and 
\be
\pa^2_\rho \varphi + c e^{-2\varphi}+ 6 e^{-2\nu-2\varphi} - Q^2 e^{-6\nu-2\varphi} = 0. 
\end{equation}

\section{Holographic dual gauge theories}  \label{holography}

 The gauge/gravity holographic duality relates a gravitational
background  in ten dimensions to  a gauge  theory in the large N limit 
residing on the boundary of the background  space-time
~\cite{Maldacena:1998re} ~\cite{Witten:1998qj} ~\cite{Aharony:1999ti}. 
It is believed that the full 
boundary gauge theory, namely,   not only its   planar limit is dual to the full string 
theory. For backgrounds with a constant dilaton, like in the
$AdS_5\times S^5$ case, 
the corresponding boundary field theory is  a CFT whereas
 backgrounds with a non-constant dilaton map into 
non-conformal boundary field theories. 
It was  conjectured in ~\cite{Polyakov:1998ju} that this concept holds also for non-critical 
SUGRA backgrounds  and even for non-supersymmetric  backgrounds that include gravity. 
In this section we explore the holographic duality of the non-critical
solutions described in the previous sections. 

Before dwelling into  the conjectured dual theories    let us first develop some intuition 
from  brane configurations that associate with  the SUGRA backgrounds we discuss.
In critical dimensions a useful way to understand the $D_p$ brane
SUGRA backgrounds ~\cite{Itzhaki:1998dd} 
is as follows. Consider
first a space-time with flat metric, constant dilaton and no
additional forms. Now place a stack of $N$ $D_p$ branes. 
The back-reaction of the latter transforms the background into that of
$D_p$ branes, namely, a curved space-time with a non-constant dilaton (apart from the $p=3$ case) 
and a $p+2$ RR form with a flux which is equal to $N$. Upon taking the near horizon limit one ends up 
in backgrounds like  for instance 
the  $AdS_5\times S^5$ for $p=3$.
In non-critical backgrounds one obviously cannot start 
with a flat $d\neq d_{critical}$ dimensional space time, and a
constant dilaton and no forms. Instead one starts with a flat $d$
dimensional Minkowski space-time with a linear dilaton  which was
shown in (\ref{eq:LinearDilaton}) 
to be a solution of the equations of motions.
The back-reaction of adding $N$ $D_p$ branes generates the  $AdS_{p+2}\times S^{d-p-2}$ backgrounds with
  $p+2$ RR forms which again have $N$ units of   flux. However, unlike the critical cases, here the dilaton
is constant for any $p$.

 A  class of backgrounds with non-constant dilaton were described in Section \ref{RRdef} associated with  
 manifolds that include an $S^1$ factor. These backgrounds can also be thought of the back-reaction of $N$
$D_p$ branes placed in manifolds of $R^{1,p+1}\times S^1$ geometry 
with a linear dilaton which is also a solution (\ref{eq:LinearDilaton}).
Recall that this background is equivalent according to 
~\cite{Giveon:1999zm},~\cite{Giveon:1999px},~\cite{Giveon:1999tq}
to the ${\cal N}=2$ super Liuville theory which is the world sheet  description
of non-critical strings.

We now proceed to describe the basic properties of the holographic
dual field theories. We start with features that are shared by 
 all the different classes of background solutions and then we
 describe each class separately.


\subsection{ The entropy  and the duality to  gauge degrees of freedom}

The basic concept of holography is that the physics of a bulk space that includes gravity
can be described by a field theory that lives on the boundary of this space.
If the boundary field theory is a $U(N)$ or $SU(N)$ gauge theory, the number
its  degrees of freedom  at the UV regime  scales like $N^2$.
To be more precise one has to introduce a UV cutoff $\delta$ and the entropy
of the gauge theory scales like

\be
S_{\textrm{gauge}}\sim  \frac{N^2}{ \delta^3},
\end{equation}
where the factor of $\frac {1}{\delta^3}$ is the number of cells 
one has to divide the volume of the three sphere on which the field theory is defined. 

Thus to verify the holographic duality one has to show that the entropy 
of the SUGRA backgrounds, that are claimed to be duals  of gauge theories, 
scale in the same manner with $N$ and $\delta$.
  To  estimate the entropy of the SUGRA bulk theory we use  
the Bekenstein-Hawking bound which is  the area in Planck units,
namely,  $S_{\textrm{SUGRA}}= \frac{\textrm{Area}}{4G_N}$.
One way to evaluate this bound is to determine the dependence of the
area and  $G_N$  on $N$ in the  string frame. Since 
in this frame the metric is $N$ independent, so is the area. On the other hand
\be 
G_N\sim e^{2\phi}\sim N^{-2} \ \ \  \rt \qquad S_{\textrm{SUGRA}}\sim N^2.
\end{equation}
Thus indeed the entropy scales as $N^2$.
Another way is to compute the area in units of $G_N$.
 The area of the boundary diverges and similarly to the field theory calculation a cutoff has to be 
introduced. For instance if one uses the following parameterization of an AdS space

\be
ds^2 = \frac{d x_{\|}^2 + dz^2}{z^2},
\end{equation} 
then the boundary is at $z=0$. Introducing a cutoff implies that the boundary is taken at $z\sim \delta$.
To compute the area in Planck units  we have to switch from
the string frame which we have used so far to the Einstein frame. The two frames are related through

\be
g^{(E)}_{ij}= e^{- \frac{4}{d-2} \phi} \ g^{(s)}_{ij}.
\end{equation}
Thus since $e^\phi\sim \frac {1}{N}$ (\ref{eq:phi0nu0})
it follows that in the Einstein frame the radii of the $AdS_{n+1}$
and the $S^k$ part  are  proportional to $N^{\frac{2}{d-2}}$, so
that the area relevant for the calculation of the entropy  bound goes like:

\bea    \label{entropygauge}
S_{\textrm{SUGRA}} &\sim& \textrm{Area} \sim V_{S^k} \left( \frac{R_{AdS}}{\delta} \right)^{n-1}
\sim \left( R_{S^k} \right)^k \left( \frac{R_{AdS}}{\delta} \right)^{n-1}
\nonumber \\
&\sim &  N^{\frac{2}{d-2} (k+n-1)} \delta^{-(n-1)}
\sim N^2  \delta^{-(n-1)},
\eea
where $\delta$ is the UV cutoff and we have used the fact that $d=n+1+k$.  
In particular in four dimensions with $n=4$ 
we find an agreement with $S_{\textrm{gauge}}$.
Hence we have shown that indeed the bound on the entropy of the SUGRA theory scales in the same way
as the boundary gauge theory. 

So far we have addressed the conformal backgrounds described in
Section (\ref{Conformal}).  
What about the RR deformed cylinder backgrounds of Section \ref{RRdef}?
It turns out that the same result applies also to that case.  
Recall, (\ref{eq:phi0nu0}) that for this case too the dilaton 
scales like $e^\phi\sim \frac {1}{N}$ and hence again 
the radii are proportional to  $N^{\frac{2}{d-2}}$ and  thus
(\ref{entropygauge}) 
holds also for the RR deformed cylinder backgrounds.


\subsection { A novel large $N$ limit}
\label{AnovelLimit}

In the original AdS/CFT duality in ten dimensions, the duality maps the bulk physics into
that of a boundary gauge theory with large $N$ and $g_s N\sim g_{YM}^2
N \gg 1$. The latter
condition follows from the requirement to have a  small scalar curvature of the background
since otherwise higher order $\alpha'$ corrections are not negligible.  
This large $N$ limit is clearly different form the perturbative one where
$N$ is taken to infinity such that $g_{YM}^2N$ is finite and small:  $g_{YM}^2N <1$.

Let us now find out  what is the consistent region of the gauge theories associated with the 
conformal backgrounds discussed in Section \ref{Conformal} 
and the RR deformed cylindrical backgrounds of  Section \ref{RRdef}.
Again the SUGRA approximation holds only provided that the scalar curvature is small.
According to (\ref{eq:AdSSRadii}) the radii of the $AdS_{n+1}$ and of the $S^k$ are  $Q$ (and hence $N$)
independent constants of order unity and so that the  curvature is 

\be
\alpha^\prime \mathcal{R}= c.
\end{equation}

Hence unlike the critical AdS/CFT duality, here 
the curvature is fixed, of order unity and cannot be reduced by taking a large $N$ limit. 

There is yet another difference between the non-critical conformal solutions and the 
$AdS_5\times S^5$ case.  
In both cases the SUGRA approximation is valid only provided the string coupling is
small. However, whereas 
in the latter case the dilaton is constant  independent of $N$,  
in the non-critical case (\ref{eq:phi0nu0}) the string coupling 
\be
g_s\sim e^{ \phi_0} = \left[\frac{1}{n+1-k} \left( \frac{(n+1-k)(k-1)}{c} \right)^k\frac{2c}{Q^2}\right ]^{1/2}
\end{equation}
and therefore  small string coupling means large $N$. 
Moreover, if we adopt the conventional correspondence between $g_s$ and $g_{YM}^2$ then 
since $g_s\sim \frac{1}{N}$, we find that the 't Hooft coupling
(\ref{eq:tHooft}) is a constant of order unity 
\be
g_{YM}^2 N \sim \left(  \frac{2 c}{n+1-k} \left( \frac{(n+1-k)(k-1)}{c} \right)^k \right)^{1/2}.
\end{equation}

To summarize, the large $N$ limit that one has to take in the  boundary gauge theory
dual to the non-critical SUGRA is different than the one taken in the critical case:

\bea
\textrm{critical}  &:&  \ \ \ N \rt \infty, \ \ \ g_{YM}^2 N \gg 1
\nonumber \\
\textrm{non-critical} &:&  \ \ \ N \rt \infty, \ \ \ g_{YM}^2 N \sim 1.
\eea  

It is interesting to note that the large $N$ limit that one has to 
take in the gauge theories which are the duals of the non-critical
SUGRAs is in between the perturbative large $N$ limit with  $g_{YM}^2 N < 1$ and the limit
in the critical cases which is $g_{YM}^2 N \gg 1$.


\subsection{ The gauge theories duals of the  $AdS_{p+2}\times S^{d-p-2}$ SUGRA backgrounds}

\label{TheGaugeTheories}

The dual field theories associated with $AdS_{p+2}\times S^{d-p-2}$ should be,
following the discussion in subsection 11.1,  conformal or
superconformal  gauge theories. 
The conformal invariance  is obviously due to 
the constant dilaton that maps into a scale invariant gauge coupling.
As for supersymmetry,  
the background solutions do solve the BPS equations.
However,  as was mentioned above this last fact is not a proof
of supersymmetry but it is a necessary condition. It is also important to note that 
the dual  gauge theories are at a fixed points in the strong 't Hooft coupling regime
since as discussed above the SUGRA corresponds to the gauge theory with $g_{YM}^2 N\sim 1$.

According to ~\cite{Kutasov:1990ua} a necessary condition for eliminating the
tachyons of  the non-critical strings is if they admit 
space-time supersymmetry which can occur only if they reside in even
space-time dimensions.
  Adopting this condition to our non-critical models, means that the dual gauge theories
can live on boundaries of bulk space-time of 
 2,4,6 and 8 dimensions. 
The gauge theories have global symmetries associated with the
isometries of the various backgrounds.
In the following table  we list in for each world volume dimensions 
the possible SUGRA models  and their corresponding global symmetries.

\TABLE[p]{
\begin{tabular}[h]{|l|l|l|}   
\hline   
 Gauge theory  & The SUGRA  & The global \\  
in $n$ dimensions & manifold &  symmetry \\
\hline    
\hline    
\ \ 2 & $AdS_3\times S^5$ & $SO(6)$ \\   
\hline
\ \ 3 & $AdS_4 $  &  -  \\   
\hline
\ \ 3 & $AdS_4\times S^2$ &  $SO(3)$ \\   
\hline
\ \ 4 & $AdS_5\times S^3$ & $SO(4)$ \\   
\hline
\ \ 5 & $AdS_6$  &  - \\   
\hline
\ \ 5 & $AdS_6\times S^2$ &  $SO(3)$ \\   
\hline
\ \ 7 & $AdS_8$ &  - \\   
\hline\hline

\end{tabular}   
\caption{The various SUGRA backgrounds and their dual gauge theories.}
}
If due to a yet unknown mechanism  the theories with odd $d$ can also
be 
stabilized then there there are additional possible dual gauge fields like the 
$AdS_5$ and the associated four dimensional gauge theory with no
global symmetry.

In order to identify  the gauge theories given in the table
with known superconformal gauge theories one has to do certain additional
checks that we leave for future investigation. In particular one should determine the 
amount of supersymmetry in the the various gravitational backgrounds.
From a brief glance on the table it seems that the two models with $SO(3)$ isometry
may correspond to superconformal gauge theories. It is known ~\cite{Minwalla:1998ka} that 
in three dimensions a theory with ${\cal N}$ supersymmetries has an R symmetry of $SO(N)$.
Hence the $AdS_4\times S^2$ model may correspond to  ${\cal N}=3$ in three dimensions.
There is a five dimensional superconformal theory with $SP(1)$ R symmetry. This may 
relate to the $AdS_6\times S^2$ model. 

For the rest of the models we
 cannot relate the data  given in the table
with known superconformal gauge theories. 
There are several logical explanations 
 to this situation:  (i) It might be that incorporating
higher curvature correction will change for instance the isometry of the transverse
manifold and hence the global symmetry of the gauge theories. 
(ii)  It might be that the dual gauge theories are
non-supersymmetric theories. For instance, one could imagine four dimensional 
theories with
four additional matter fields in the adjoint that admit the  $SO(4)$ global
symmetry and strongly coupled fixed points.
(iii) It might be that only part of the full isometry translates 
into an R symmetry or a global symmetry of the gauge theory due to the
fact, that the GSO projection is
compatible only with a subgroup of the full isometry group. 
Such a case occurs for the $AdS_3\times S^3$ ~\cite{Giveon:1999jg}.
One may imagine a situation where
only the subgroup $SU(2)\times U(1)$ of the $SO(4)$ isometry group 
survives the GSO projection and hence a potential 
compatibility with the $R$ symmetry of ${\cal N}=2$.

Assuming that there are conformal gauge theories that correspond 
to these non-critical SUGRA backgrounds, one can turn on the known
machinery of computing 
the conformal dimensions of chiral operators computing correlation
functions etc. in a similar manner to what was done in the critical cases.


\subsection{ The AdS black hole solutions and their  gauge theory
  duals} 
\label{AdSBHandDualGT}

Les us now examine  the gauge theory duals of the 
 AdS black hole solutions discussed in Subsection \ref{AdSBH}.
 Note that the latter  have a thermal factor in the metric that
resembles 
 that of  near extremal critical $D_p$ branes ~\cite{Itzhaki:1998dd}.
However whereas in the former case the thermal factor is 
$1-(\frac{u_0}{u})^n= 1-(\frac{u_0}{u})^{p+1}$ in the later case
it is $1-(\frac{u_0}{u})^{7-p}$.
Remarkably in  the four dimensional case  ($n=4, p=3$) the 
non-critical $AdS_5$ black hole solution matches that of the black hole of
$AdS_5\times S^5$. (We refer to the AdS part, obviously they have
different $S^k$). However there is still one difference between the two cases
and that is the  the AdS
radius:
\bea
\textrm{critical}      & : &  R_{AdS} = \left( g_{YM}\sqrt{N} \right)^{1/2}  \\
\textrm{non-critical}  & : &  R_{AdS} = \left( \frac{n(n+1-k)}{c} \right)^{1/2}.
\eea
Due to the similarity between the critical and non-critical near
extremal solutions, 
we do not have to redo the calculations in the gauge theory but rather read them from the known 
results of the near extremal $AdS_5\times S^5$. solution. 
In particular, we can implement the idea proposed in ~\cite{Witten:1998zw} of 
imposing anti-periodic boundary conditions while taking the
large temperature limit, namely small thermal radius of the background, which leads to 
a pure YM theory in three dimensions. 
The properties of the 3d
gauge dynamics that we will  focus on are the Wilson loop and the glue-ball spectrum.

To determine the Wilson
loop one can write down the NG action associated with the background
metric and determine the classical configuration of the
string ~\cite{Brandhuber:1998bs} ~\cite{Brandhuber:1998er}. 
Instead we can use the general result of ~\cite{Kinar:1998pj}
~\cite{Kinar:1998vq} ~\cite{Sonnenschein:1999yb}. According
to these results if one of the following two conditions is obeyed, the
corresponding Wilson line admits an area law behavior:
\bea
 g_{00}g_{ii}(\tau) \quad  \textrm{has  a  minimum  at} \quad \tau_{\textrm{min}}  \quad
 \textrm{with} \quad  
 g_{00}g_{ii} (\tau_{\textrm{min}}) & > & 0,\\ \nonumber 
 g_{00}g_{\tau\tau} (\tau) \quad \textrm{diverges at} \quad   \tau_{\textrm{div}}
 \quad \textrm{with} \quad
 g_{00}g_{ii}(\tau_{\textrm{div}})  & > & 0. \\ \nonumber
\eea
It is easy to check that after the reduction to 3d $g_{00}g_{uu}= [1-(\frac{u_0}{u})^4]^{-1}$ and it 
 diverges at $u=u_0$. The conclusion is therefore that indeed the 
Wilson loop in this background admits an area law behavior with string tension of
\be
\textrm{string tension} = \frac{1}{2\pi} \left(\frac {u_0}{R_{AdS}}\right)^2 
       =  \frac{1}{2}\pi \frac{8}{c} T^2,
\end{equation}
where $T$ is the temperature which is related to $u_0$ as follows $u_0= \pi R^2_{AdS} T$.
It is interesting to compare this result to the one derived from the critical AsS black hole.
In the latter case the factor  $\frac{8}{c}$ is replaced by $\sqrt
{g_{YM}^2 N}$
 so that one has later to interpolate between this result and the one
 associated with finite ${g_{YM}^2 N}$. In our case this last step is
 obviously unnecessary. 

To describe the four dimensional Wilson loop, one can start  ~\cite{Witten:1998zw}
with the conformal background of $M^5$ branes, namely $AdS_7\times S^4$, introduce two circles
and in the limit of small radii the world volume manifold is reduced from a six dimensional one to 
a four dimensions. Alternatively, one can take   
the infinite temperature limit of the non-conformal $D_4$ SUGRA
background ~\cite{Brandhuber:1998bs} ~\cite{Brandhuber:1998er}.
Here using the non-critical solutions we can start with the near extremal
$AdS_6$ solution, introduce temperature, and take the infinite temperature limit thus getting
a four dimensional theory. 
The outcome of the calculation  of the string tension
 will be the same as above
(changing the factor 8 into 30). 
Note that there is a difference between that and the critical case
where the
 4d string tension is  $\frac{1}{2\pi} (\frac {u_0}{R_{AdS}})^{3/2}$.
 
Next we consider the glue-ball spectrum.
The analysis of the three dimensional glue-balls is identical to the one done in the near extremal 
$AdS_5\times S^5$ background. Solving the equation of motion of the dilaton in the non-critical near 
extremal background, choosing the normalizable solution, imposing regularity at the horizon, one finds
a discrete spectrum with a mass gap.
For instance the masses that correspond to the  $0^{++}$ glue-balls are
$M^2_{0^{++}}\sim \frac{1}{T^2}$. 
Since this result is independent of $R_{AdS}$ it is identical to the one in the critical case.
However there is a physical difference between the picture in the non-critical and critical cases 
since in the latter case the string tension is proportional to
$\sqrt{g_{YM}^2 N}$,
$M^2_{0^{++}}$ does not scale like the string tension, whereas in our case it does.

To get the glue-ball spectrum in four dimensions one has to solve the equation of motion of the dilaton
in the background achieved from that of the  near extremal $AdS_6$ in the zero radius limit.
Assuming the fluctuation of the dilaton is
$\delta\phi=\tilde\phi(u) e^{ikx}$,  the Laplacian that enters 
the equation, which determines the glue-ball spectrum  
\be
\pa_u[u(u^5-u_0^5)\pa_u]\tilde\phi(u) + M^2u\tilde\phi(u), 
\qquad \textrm{where} \qquad
M^2 =-k^2
\end{equation}
 is different than the one solved in the critical case.

\subsection{ The gauge duals of the RR deformed cylinder SUGRA
  backgrounds}

\label{DualRRdef}

The RR deformed cylindrical solutions were found for any world-volume dimension $n$.
Let us concentrate on the four dimensional case. 
As discussed above from the entropy of the SUGRA it seems natural that the 
boundary gauge theory has an $SU(N)$ gauge symmetry.
Since the dilaton is varying the dual gauge theory is non-conformal.
As was stated above we have not analyzed the amount of supersymmetry
these backgrounds admit. However if we relate to the naive brane
configuration constructed on the $R^{1,3}\times cigar$ we would
conclude that the 
 dual four dimensional gauge theory is ${\cal N}=1$ supersymmetric.
By construction these solutions have an $SO(2)$ isometry associated
with the $S^1$ part
 of the manifold and hence one expects that the dual gauge theory has an $U_R(1)$ symmetry. 
Combining these three ingredients of the local symmetry group, the
supersymmetry and the R symmetry,
 indicates that the dual gauge theory maybe a close cousin of the  ${\cal N}=1$ SYM theory. 
One clear difference  this theory and the dual of our SUGRA
backgrounds 
is the $U_R(1)$ symmetry  since in our SUGRA it is a unbroken symmetry
whereas in ${\cal N}=1$ SYM theory it is  broken both by instantons
and also spontaneously. 
To look for backgrounds that correspond to a spontaneously broken
symmetry one has to abandon our original
ansatz where all the fields depend only on the radial direction and
not on the $\theta$ (the coordinate along the $S^1$) direction. Modifying the ansatz will
render the task of finding solutions to BPS equations much more
complicated so we 
left it for future investigation.

If indeed the dual gauge theory is a relative of the   ${\cal N}=1$
SYM theory, it should admit confinement. As was discussed above this can be checked both by
the determination  of the expectation value of the Wilson loop as well
as by  a computation of the glue-ball spectrum. 
Using the necessary conditions to have confining Wilson loop given above
it is easy to see that 
 the first condition translates into having a minimum to
$e^\lambda$ at a non vanishing value.  
A convenient way to check if this condition is obeyed is to examine
Fig.\ref{e2L}.

It is clear that apart from the class of solutions described by the
line that ends on point A the rest of the solutions do admit a minimum for  
\be
\partial_\tau e^\lambda(\tau_{\textrm{min}})=0 
 \qquad  e^\lambda(\tau_{\textrm{min}})>0 .    
\end{equation}
Thus according to the above criterion the gauge theory dual to the RR
deformed SUGRA backgrounds indeed admits confinement behavior. 
Note however that unlike in critical duals of confining gauge theories
that are close relatives of ${\cal N}=1$
SYM theory like those of (~\cite{Klebanov:2000hb},
~\cite{Maldacena:2000yy}), in the present case the
minimum of $e^\lambda$ does not occur at the ``end'' of the radial
direction  but rather at some mid point which means that in fact the
bulk space-time has two boundaries. Backgrounds of this form were
found also in ~\cite{Armoni:1999fb}. Since beyond $\tau_{min}$ the validity of
the SUGRA approximation is doubtful we will not attempt to discuss 
the background at that region. Generically if indeed there are two
boundaries one has to be worried about loss of unitarity on the
boundary field theory due to ``leakage'' of signals to the other
boundary.
As was mentioned above a background which is a faithful dual of the     
 ${\cal N}=1$ SYM theory, has to admit a spontaneous $U_R(1)$ breaking
 as well as a breaking due to instantons. These phenomena cannot take
 place using our ansatz of dependence only on the radial coordinate.
One may anticipate that solutions that share a breaking of the
$U_R(1)$ will have a different radial dependence so that the problem
of two boundaries will be avoided.

\section{ Summary and discussion}

There are several reasons to believe that the string theory of QCD is not a ten dimensional theory. 
KK states that cannot be decoupled from the physical hadronic spectrum is one such a phenomenon.
This situation calls for string theory in less than ten dimension.
Due to our ignorance about quantizing string theories on RR
backgrounds, in the NSR formulation 
we cannot do much in the study
of non-critical string theories that may be candidates for describing hadronic physics. This pushes us
to address first the small $\alpha'$ limit of string theories, namely the low energy SUGRA solutions.
That was precisely the aim of this paper.

The equations of motion that follow from the requirement for vanishing of the  $\beta$ function
on the world sheet of the string, are in general a set of complicated coupled second order partial 
differential equations. The equations for non-critical strings are even more complicated due to an
additional term in the potential. To simplify the system we choose a particular ansatz of the metric
which is geared toward holographic applications. This ansatz assumes 
dependence only on a  radial direction thus turning the problem into 
that of ordinary differential equations.
A further technical simplification  is achieved by converting the
problem into BPS first order differential equations using the superpotential formalism. 
From all possible topologies we have focused on solution of two types:  backgrounds 
with a structure of  $AdS_{p+2}\times S^{d-p-2}$ and those that asymptote a linear dilaton with 
a topology of $R^{1,d-3}\times R \times S^1$. Whereas the former class are analytic solutions the latter
include numerical solutions with approximated analytic behavior close to the origin.

The SUGRA solutions we have found can serve as an anti holographic description of gauge theories in
a particular large $N$ limit which is neither the perturbative 
nor that of the critical AdS correspondence. 
It is characterized by $N\rt \infty$ and $g_{YM}^2 N \sim 1$.
We have made the first steps of the analysis of the properties which
are of interest from 
the gauge dynamics point of view like the Wilson loops and the glue-ball spectra.

Let us now propose several open questions that deserve  further
investigation:

\begin{itemize}

\item
 A very essential issue that we  have not addressed in this
paper is controlling  the higher curvature corrections. Generically we
do not have yet any evidence that the latter are small and hence it
will be interesting to compute the impact of the leading order
curvature correction on the backgrounds discussed. Our belief is that 
the general structure of the background  space-time will not be changed
but the characterizing parameters will be modified, that is to say
that for instance the $AdS_{n+1}\times S^k$ structure will remain but
the radii of the $AdS_{n+1}$ and $S^k$ will be modified. 
\item
In the search of solutions of the equation of motions of the
non-critical low energy effective action (\ref{eq:TheAction}) we did not take the 
most general ansatz both for the metric and for the various forms.
One can extend our search  by looking for 
(i) more general metric in particular other possible transverse spaces
like the conifold,  
(ii)   solutions with NS form combined with a RR form,
(iii)  solutions with more than one RR form,
(iv) solutions where the metric, dilaton and the form depend on other
coordinates rather only the radial one. 
\item
The full analysis of the properties of the holographic dual gauge
theories has to be performed. In particular determining the glue-ball
spectrum quantum corrections of the Wilson loop, Baryons etc. 
(see ~\cite{Aharony:2002up},~\cite{Loewy:2001pq}) in the
context of the confining backgrounds like the AdS black hole solutions
and the RR deformed cylindrical solutions.
\item
One established route to  go beyond the SUGRA limit is to invoke the
Penrose limit and quantize the plane wave string theory. 
Hopefully it can be done for the class of the $AdS_{n+1}\times S^k$
backgrounds and the corresponding black hole solutions. An interesting
challenge is to apply the limit also for the solutions that were found
only numerically. 
\item
To incorporate dynamical quark in the dual gauge theories, one may
use the idea of introducing probes similar to the probes introduced
for confining critical backgrounds ~\cite{Sakai:2003wu} and deducing 
the mesonic spectrum from the spectrum of the probes fluctuations.

\end{itemize}

\acknowledgments

We would like to thank S. Elitzur, A. Giveon, N. Itzhaki, D. Kutasov,
Y. Oz and T. Sakai  for fruitful discussions.
We would  specially like to  thank Ofer Aharony for many useful
conversations about the project and for
his illuminating  comments about the manuscript. 
J.S would like to thank O. Ganor, I. Klebanov and J. Maldacena for discussions
at  the very  early stage of the project. 
This work was supported in part by the German-Israeli Foundation for
Scientific Research and by the Israel Science Foundation.

\appendix

\section{The derivation of the quantum mechanical action} 
\label{ThaActionDerivation}

For the metric
$$
l_s^{-2} ds^2 = d\tau^2 +e^{2\lambda} dx_{\|}^2 + e^{2\nu} d\Omega_k^2,
$$
we calculate 
\bea
\alpha^\prime \mathcal{R} &=&
-2n\lambda'' -n(n+1)(\lambda')^2
+k(k-1)e^{-2\nu}
\non &&
-2k\nu'' 
-2k n\nu'\lambda'
-k(k+1)(\nu')^2,
\non
\sqrt{G} &=& e^{n\lambda+k\nu}.
\non
\eea
Thus
\bea
S &=& \int\sqrt{G} \mathcal{R} =
{\alpha^\prime}^{-1} \int
e^{n\lambda+k\nu}\lbrack
(2n\lambda' +2k\nu')(n\lambda'+k\nu') -n(n+1)(\lambda')^2
\non &&
+k(k-1)e^{-2\nu}-2k n\nu'\lambda'
-k(k+1)(\nu')^2
\rbrack
\non &=&
{\alpha^\prime}^{-1} \int
e^{n\lambda+k\nu}\lbrack
n(n-1)(\lambda')^2
+k(k-1)e^{-2\nu}+2k n\nu'\lambda'
+k(k-1)(\nu')^2
\rbrack
\non
&=&
{\alpha^\prime}^{-1} \int
e^{n\lambda+k\nu}\lbrack
(n\lambda' + k\nu')^2
-n(\lambda')^2
+k(k-1)e^{-2\nu}
-k(\nu')^2
\rbrack.
\non
\eea

\section{From the BPS equations to the equations of motion}

\label{FromBPSToEquations}

The BPS equations take the form
\be
{f_a}'= \half G_{ab}\pa^b W 
\end{equation}
Differentiating  with respect to $\tau$ we get
\bea
{f_a}''&=& \half [ \pa_cG_{ab}{f^c}'\pa^b W +  G_{ab}\pa_c\pa^b W {f^c}']\CR
 &=& \pa_cG_{ab}{f^c}'{f^b}'+ \frac{1}{4} G^{cd}\pa_a \pa_c W\pa_d W\CR
&=& \pa_cG_{ab}{f^c}'{f^b}' + \frac{1}{8} \pa_a (G^{cd} \pa_c W\pa_d W)- \frac{1}{8}\pa_a (G^{cd}
\pa_c W\pa_d W) \CR
&=&\pa_cG_{ab}{f^c}'{f^b}' + \pa_a V - \frac{1}{2} \pa_a G^{bc}{f_c}'{f_b}'\CR
&=&\pa_cG_{ab}{f^c}'{f^b}' + \pa_a V + \frac{1}{2} \pa_a G_{bc}{f^c}'{f^b}'\CR
\eea
and the last line is indeed the equation of motion (\ref{seqom}).


\section{Action and BPS equations in Einstein metric}

\label{EinsteinFrame}

Define the Einstein metric
\be
ds^2_E= e^{-\frac{4}{n+k-1}\phi} ds^2 
\end{equation}
where $ds^2$ is the string frame metric, and parameterize the Einstein metric
for $k\neq 0$ as follow
\be
ds^2_E= e^{\frac{-2k}{n-1}B}\left(du^2 + e^{\frac{8}{n}A(u) }dx_{\|}^2 \right)
+ e^{2B(u)}d^2\Omega_k 
\end{equation}
so that the following relations hold
\be\label{relation}
\frac{4}{n+k-1}\phi - \frac{2k}{n-1}B + \frac{8}{n}A= 2\lambda; 
 \qquad \frac{4}{n+k-1}\phi+ 2B=2\nu;
\end{equation}
and 
\be
d\tau = du \, e^{\frac{4}{n+k-1}\phi  - \frac{2k}{n-1}B},
\end{equation}
then the corresponding action has the form of (\ref{actionE})
\bea
S &=& \int du  e^{4A}\left(3 (A')^2-\frac{3}{16} \frac{kn(n+k-1)}{(n-1)^2}(B')^2 
-\frac{3}{4} \frac{n}{(n-1)(n+k-1)}(\phi')^2 -V_E(f)\right) \CR
V_E &=& \frac{3}{16} \frac{n}{n-1}\left[ -e^{-\frac{2k}{n-1}B}(c e^{\frac{4}{n+k-1}\phi}+
\Lambda e^{  \frac{2(n+k+1)}{(n+k-1)}\phi})
+k(k-1)e^{- \frac{(n+k-1)}{(n-1)}B} +V_{RR}
\right]\CR
V_{RR}&=& \frac{1}{4}N^2 e^{2\phi-  \frac{2(n+k-1)}{ (n-1)}B},
\eea
which implies that  $G^{\phi\phi}= \frac{2}{3} \frac{(n-1)(n+k-1)}{n}$ and 
 $G^{BB}= \frac{8}{3}\frac{(n-1)^2}{kn(n+k-1)}$.

\section{The cigar solution in the Einstein frame}

\label{TheCigarEinsteinframe}

Another superpotential that corresponds to the potential 
\be\label{VE}
 V_E =  - \frac{3}{16} c \frac{n}{(n-1)} e^{(\frac{4}{n}\phi-\frac{2}{n-1}B)}
\end{equation}
 takes the form
\be
W_E = -\frac{3}{8} \frac{n}{n-1} \sqrt{c}\left[ e^{\frac{4}{n}\phi  + \frac{n-2}{n-1}B} 
+  e^{-\frac{n}{n-1}B} \right]
\end{equation}
 Defining now a new radial coordinate $dr=2 \sqrt{c} e^{\frac{2}{n}\phi-\frac{1}{n-1}B}du $
 the BPS equations read
\bea
{\pa_r B} &=& -\left[\frac{(n-2)}{n} e^{\frac{2}{n}\phi +B} -  e^{-\frac{2}{n}\phi -B} \right]\CR
{\pa_r \phi} &=& - \left[ e^{\frac{2}{n}\phi + B}\right]\CR
{\pa_r A}  &=& \frac{1}{8} \frac{n}{n-1} \left[ e^{\frac{2}{n}\phi +B} +  e^{-\frac{2}{n}\phi -B} \right]\CR
\eea

These equations have the following solution 
\bea
&& \phi = -\ln(\cosh(r)), \qquad  B=\ln( \sinh(r)) - \frac{n-2}{n}\ln(\cosh(r)) \CR
&&A=\frac{n}{4(n-1)} \ln(\cosh(r)\sinh(r)),
\eea
which means that 
\be
 e^{\frac{4}{n}\phi +2B}= \tanh^2r  \qquad  e^{\frac{4}{n}\phi - \frac{2}{n-1}B +\frac{8}{n}A} =1,
\end{equation}
so that the metric is that of a cigar, namely:
\be
ds^2 =  dr^2  +  dx_{\|}^2 + \tanh^2(r)   d^2\theta.  
\end{equation}


\section{The BGV solution from the equations of motion}

 \label{BGVsolution}

Using the facts that $d\tau=\frac{d \phi}{\sqrt{l(\phi)}}$ and 
$d\tau=e^{-\varphi} d \rho$ we can re-write
the equations of motion (\ref{EMrhophi}) for $n=1$ and $k=0$ 
in the formulation, where the dilaton is the radial direction:
\be
\pa_\phi^2 \lambda +  2\pa_\phi \lambda(\pa_\phi \lambda -1) 
   - \frac{Q^2}{2}e^{-2(\lambda-\phi)}=0
\end{equation}
and 
\be
 (\pa_\phi \lambda -1) + \frac{c}{4}e^{-2\lambda}
      -\frac{Q^2}{2}e^{-2(\lambda-\phi)}=0.
\end{equation}
The solution to the first equation
can be obtained as follows.
First we write:
\bea
\pa_\phi^2 \left( e^{2\lambda-\phi} \right) &=&
2e^{2\lambda-\phi}(\pa_\phi^2\lambda +\frac{1}{2}(2\pa_\phi\lambda -1)^2)
\non &=&
2e^{2\lambda-\phi} \left(\pa_\phi^2\lambda +2\pa_\phi\lambda(\pa_\phi\lambda -1) +\frac{1}{2} \right)
\non &=&
-Q^2 e^{\phi} + e^{2\lambda-\phi}.
\nonumber
\eea
Setting $\eta\equiv e^{2\lambda-\phi}$ we obtain
$$
\eta'' - \eta = -Q^2 e^{\phi}
$$
and the solution is
$$
\eta = - \frac{Q^2}{2}(\phi +C_0) e^{\phi} + C_1 e^{-\phi}.
$$
Thus
$$
e^{2\lambda} = - \frac{Q^2}{2}(\phi +C_0) e^{2\phi} + C_1.
$$
Finally, substituting this solution into the dilaton equation we
obtain that $C_1=\frac{c}{4}$.

\bibliography{ncsg}

\providecommand{\href}[2]{#2}\begingroup\raggedright\begin{thebibliography}{10}

\bibitem{Kutasov:1990ua}
D.~Kutasov and N.~Seiberg, {\it Noncritical superstrings},  {\em Phys. Lett.}
  {\bf B251} (1990) 67--72.

\bibitem{Kutasov:1991pv}
D.~Kutasov, {\it Some properties of (non)critical strings},
  \href{http://xxx.lanl.gov/abs/hep-th/9110041}{{\tt hep-th/9110041}}.

\bibitem{Giveon:1999zm}
A.~Giveon, D.~Kutasov, and O.~Pelc, {\it Holography for non-critical
  superstrings},  {\em JHEP} {\bf 10} (1999) 035,
  [\href{http://xxx.lanl.gov/abs/hep-th/9907178}{{\tt hep-th/9907178}}].

\bibitem{Giveon:1999px}
A.~Giveon and D.~Kutasov, {\it Little string theory in a double scaling limit},
   {\em JHEP} {\bf 10} (1999) 034,
  [\href{http://xxx.lanl.gov/abs/hep-th/9909110}{{\tt hep-th/9909110}}].

\bibitem{Giveon:1999tq}
A.~Giveon and D.~Kutasov, {\it Comments on double scaled little string theory},
   {\em JHEP} {\bf 01} (2000) 023,
  [\href{http://xxx.lanl.gov/abs/hep-th/9911039}{{\tt hep-th/9911039}}].

\bibitem{Hori:2001ax}
K.~Hori and A.~Kapustin, {\it Duality of the fermionic 2d black hole and n = 2
  liouville theory as mirror symmetry},  {\em JHEP} {\bf 08} (2001) 045,
  [\href{http://xxx.lanl.gov/abs/hep-th/0104202}{{\tt hep-th/0104202}}].

\bibitem{Polyakov:1998ju}
A.~M. Polyakov, {\it The wall of the cave},  {\em Int. J. Mod. Phys.} {\bf A14}
  (1999) 645--658, [\href{http://xxx.lanl.gov/abs/hep-th/9809057}{{\tt
  hep-th/9809057}}].

\bibitem{Klebanov:1998yy}
I.~R. Klebanov and A.~A. Tseytlin, {\it D-branes and dual gauge theories in
  type 0 strings},  {\em Nucl. Phys.} {\bf B546} (1999) 155--181,
  [\href{http://xxx.lanl.gov/abs/hep-th/9811035}{{\tt hep-th/9811035}}].

\bibitem{Ferretti:1998xu}
G.~Ferretti and D.~Martelli, {\it On the construction of gauge theories from
  non critical type 0 strings},  {\em Adv. Theor. Math. Phys.} {\bf 3} (1999)
  119--130, [\href{http://xxx.lanl.gov/abs/hep-th/9811208}{{\tt
  hep-th/9811208}}].

\bibitem{Ferretti:1999gj}
G.~Ferretti, J.~Kalkkinen, and D.~Martelli, {\it Non-critical type 0 string
  theories and their field theory duals},  {\em Nucl. Phys.} {\bf B555} (1999)
  135--156, [\href{http://xxx.lanl.gov/abs/hep-th/9904013}{{\tt
  hep-th/9904013}}].

\bibitem{Klebanov:1998yz}
I.~R. Klebanov and A.~A. Tseytlin, {\it Asymptotic freedom and infrared
  behavior in the type 0 string approach to gauge theory},  {\em Nucl. Phys.}
  {\bf B547} (1999) 143--156,
  [\href{http://xxx.lanl.gov/abs/hep-th/9812089}{{\tt hep-th/9812089}}].

\bibitem{Garousi:1999fu}
M.~R. Garousi, {\it String scattering from d-branes in type 0 theories},  {\em
  Nucl. Phys.} {\bf B550} (1999) 225--237,
  [\href{http://xxx.lanl.gov/abs/hep-th/9901085}{{\tt hep-th/9901085}}].

\bibitem{Minahan:1999yr}
J.~A. Minahan, {\it Asymptotic freedom and confinement from type 0 string
  theory},  {\em JHEP} {\bf 04} (1999) 007,
  [\href{http://xxx.lanl.gov/abs/hep-th/9902074}{{\tt hep-th/9902074}}].

\bibitem{Nekrasov:1999mn}
N.~Nekrasov and S.~L. Shatashvili, {\it On non-supersymmetric cft in four
  dimensions},  {\em Phys. Rept.} {\bf 320} (1999) 127--129,
  [\href{http://xxx.lanl.gov/abs/hep-th/9902110}{{\tt hep-th/9902110}}].

\bibitem{Billo:1999nf}
M.~Billo, B.~Craps, and F.~Roose, {\it On d-branes in type 0 string theory},
  {\em Phys. Lett.} {\bf B457} (1999) 61--69,
  [\href{http://xxx.lanl.gov/abs/hep-th/9902196}{{\tt hep-th/9902196}}].

\bibitem{Armoni:1999fb}
A.~Armoni, E.~Fuchs, and J.~Sonnenschein, {\it Confinement in 4d yang-mills
  theories from non-critical type 0 string theory},  {\em JHEP} {\bf 06} (1999)
  027, [\href{http://xxx.lanl.gov/abs/hep-th/9903090}{{\tt hep-th/9903090}}].

\bibitem{Imamura:1999um}
Y.~Imamura, {\it Branes in type 0 / type ii duality},  {\em Prog. Theor. Phys.}
  {\bf 102} (1999) 859--866,
  [\href{http://xxx.lanl.gov/abs/hep-th/9906090}{{\tt hep-th/9906090}}].

\bibitem{Ghoroku:1999bk}
K.~Ghoroku, {\it Yang-mills theory from non-critical string},  {\em J. Phys.}
  {\bf G26} (2000) 233--244,
  [\href{http://xxx.lanl.gov/abs/hep-th/9907143}{{\tt hep-th/9907143}}].

\bibitem{Maloney:2002rr}
A.~Maloney, E.~Silverstein, and A.~Strominger, {\it De sitter space in
  noncritical string theory},
  \href{http://xxx.lanl.gov/abs/hep-th/0205316}{{\tt hep-th/0205316}}.

\bibitem{Papadopoulos:2000gj}
G.~Papadopoulos and A.~A. Tseytlin, {\it Complex geometry of conifolds and
  5-brane wrapped on 2- sphere},  {\em Class. Quant. Grav.} {\bf 18} (2001)
  1333--1354, [\href{http://xxx.lanl.gov/abs/hep-th/0012034}{{\tt
  hep-th/0012034}}].

\bibitem{Borokhov:2002fm}
V.~Borokhov and S.~S. Gubser, {\it Non-supersymmetric deformations of the dual
  of a confining gauge theory},  {\em JHEP} {\bf 05} (2003) 034,
  [\href{http://xxx.lanl.gov/abs/hep-th/0206098}{{\tt hep-th/0206098}}].

\bibitem{Kuperstein:2003yt}
S.~Kuperstein and J.~Sonnenschein, {\it Analytic non-supersymmtric background
  dual of a confining gauge theory and the corresponding plane wave theory of
  hadrons},  {\em JHEP} {\bf 02} (2004) 015,
  [\href{http://xxx.lanl.gov/abs/hep-th/0309011}{{\tt hep-th/0309011}}].

\bibitem{Giveon:1994fu}
A.~Giveon, M.~Porrati, and E.~Rabinovici, {\it Target space duality in string
  theory},  {\em Phys. Rept.} {\bf 244} (1994) 77--202,
  [\href{http://xxx.lanl.gov/abs/hep-th/9401139}{{\tt hep-th/9401139}}].

\bibitem{Berkovits:2001tg}
N.~Berkovits, S.~Gukov, and B.~C. Vallilo, {\it Superstrings in 2d backgrounds
  with r-r flux and new extremal black holes},  {\em Nucl. Phys.} {\bf B614}
  (2001) 195--232, [\href{http://xxx.lanl.gov/abs/hep-th/0107140}{{\tt
  hep-th/0107140}}].

\bibitem{Aharony:1998ub}
O.~Aharony, M.~Berkooz, D.~Kutasov, and N.~Seiberg, {\it Linear dilatons,
  ns5-branes and holography},  {\em JHEP} {\bf 10} (1998) 004,
  [\href{http://xxx.lanl.gov/abs/hep-th/9808149}{{\tt hep-th/9808149}}].

\bibitem{Kounnas:1994py}
C.~Kounnas, {\it Construction of superstrings in wormhole like backgrounds},
  \href{http://xxx.lanl.gov/abs/hep-th/9402080}{{\tt hep-th/9402080}}.

\bibitem{Myers:1987fv}
R.~C. Myers, {\it New dimensions for old strings},  {\em Phys. Lett.} {\bf
  B199} (1987) 371.

\bibitem{Alvarez:2000it}
E.~Alvarez, C.~Gomez, and L.~Hernandez, {\it Non-critical poincare invariant
  bosonic string backgrounds and closed string tachyons},  {\em Nucl. Phys.}
  {\bf B600} (2001) 185--196,
  [\href{http://xxx.lanl.gov/abs/hep-th/0011105}{{\tt hep-th/0011105}}].

\bibitem{Callan:1991at}
J.~Callan, Curtis~G., J.~A. Harvey, and A.~Strominger, {\it Supersymmetric
  string solitons},  \href{http://xxx.lanl.gov/abs/hep-th/9112030}{{\tt
  hep-th/9112030}}.

\bibitem{Murthy:2003es}
S.~Murthy, {\it Notes on non-critical superstrings in various dimensions},
  {\em JHEP} {\bf 11} (2003) 056,
  [\href{http://xxx.lanl.gov/abs/hep-th/0305197}{{\tt hep-th/0305197}}].

\bibitem{Itzhaki:1998dd}
N.~Itzhaki, J.~M. Maldacena, J.~Sonnenschein, and S.~Yankielowicz, {\it
  Supergravity and the large n limit of theories with sixteen supercharges},
  {\em Phys. Rev.} {\bf D58} (1998) 046004,
  [\href{http://xxx.lanl.gov/abs/hep-th/9802042}{{\tt hep-th/9802042}}].

\bibitem{Maldacena:1998re}
J.~M. Maldacena, {\it The large n limit of superconformal field theories and
  supergravity},  {\em Adv. Theor. Math. Phys.} {\bf 2} (1998) 231--252,
  [\href{http://xxx.lanl.gov/abs/hep-th/9711200}{{\tt hep-th/9711200}}].

\bibitem{Witten:1998qj}
E.~Witten, {\it Anti-de sitter space and holography},  {\em Adv. Theor. Math.
  Phys.} {\bf 2} (1998) 253--291,
  [\href{http://xxx.lanl.gov/abs/hep-th/9802150}{{\tt hep-th/9802150}}].

\bibitem{Aharony:1999ti}
O.~Aharony, S.~S. Gubser, J.~M. Maldacena, H.~Ooguri, and Y.~Oz, {\it Large n
  field theories, string theory and gravity},  {\em Phys. Rept.} {\bf 323}
  (2000) 183--386, [\href{http://xxx.lanl.gov/abs/hep-th/9905111}{{\tt
  hep-th/9905111}}].

\bibitem{Minwalla:1998ka}
S.~Minwalla, {\it Restrictions imposed by superconformal invariance on quantum
  field theories},  {\em Adv. Theor. Math. Phys.} {\bf 2} (1998) 781--846,
  [\href{http://xxx.lanl.gov/abs/hep-th/9712074}{{\tt hep-th/9712074}}].

\bibitem{Giveon:1999jg}
A.~Giveon and M.~Rocek, {\it Supersymmetric string vacua on ads(3) x n},  {\em
  JHEP} {\bf 04} (1999) 019,
  [\href{http://xxx.lanl.gov/abs/hep-th/9904024}{{\tt hep-th/9904024}}].

\bibitem{Witten:1998zw}
E.~Witten, {\it Anti-de sitter space, thermal phase transition, and confinement
  in gauge theories},  {\em Adv. Theor. Math. Phys.} {\bf 2} (1998) 505--532,
  [\href{http://xxx.lanl.gov/abs/hep-th/9803131}{{\tt hep-th/9803131}}].

\bibitem{Brandhuber:1998bs}
A.~Brandhuber, N.~Itzhaki, J.~Sonnenschein, and S.~Yankielowicz, {\it Wilson
  loops in the large n limit at finite temperature},  {\em Phys. Lett.} {\bf
  B434} (1998) 36--40, [\href{http://xxx.lanl.gov/abs/hep-th/9803137}{{\tt
  hep-th/9803137}}].

\bibitem{Brandhuber:1998er}
A.~Brandhuber, N.~Itzhaki, J.~Sonnenschein, and S.~Yankielowicz, {\it Wilson
  loops, confinement, and phase transitions in large n gauge theories from
  supergravity},  {\em JHEP} {\bf 06} (1998) 001,
  [\href{http://xxx.lanl.gov/abs/hep-th/9803263}{{\tt hep-th/9803263}}].

\bibitem{Kinar:1998pj}
Y.~Kinar, E.~Schreiber, and J.~Sonnenschein, {\it Precision 'measurements' of
  the q anti-q potential in m{QCD}},  {\em Nucl. Phys.} {\bf B544} (1999)
  633--649, [\href{http://xxx.lanl.gov/abs/hep-th/9809133}{{\tt
  hep-th/9809133}}].

\bibitem{Kinar:1998vq}
Y.~Kinar, E.~Schreiber, and J.~Sonnenschein, {\it Q anti-q potential from
  strings in curved spacetime: Classical results},  {\em Nucl. Phys.} {\bf
  B566} (2000) 103--125, [\href{http://xxx.lanl.gov/abs/hep-th/9811192}{{\tt
  hep-th/9811192}}].

\bibitem{Sonnenschein:1999yb}
J.~Sonnenschein, {\it Wilson loops from supergravity and string theory},  {\em
  Class. Quant. Grav.} {\bf 17} (2000) 1257--1266,
  [\href{http://xxx.lanl.gov/abs/hep-th/9910089}{{\tt hep-th/9910089}}].

\bibitem{Klebanov:2000hb}
I.~R. Klebanov and M.~J. Strassler, {\it Supergravity and a confining gauge
  theory: Duality cascades and chisb-resolution of naked singularities},  {\em
  JHEP} {\bf 08} (2000) 052,
  [\href{http://xxx.lanl.gov/abs/hep-th/0007191}{{\tt hep-th/0007191}}].

\bibitem{Maldacena:2000yy}
J.~M. Maldacena and C.~Nunez, {\it Towards the large n limit of pure n = 1
  super yang mills},  {\em Phys. Rev. Lett.} {\bf 86} (2001) 588--591,
  [\href{http://xxx.lanl.gov/abs/hep-th/0008001}{{\tt hep-th/0008001}}].

\bibitem{Aharony:2002up}
O.~Aharony, {\it The non-ads/non-cft correspondence, or three different paths
  to qcd},  \href{http://xxx.lanl.gov/abs/hep-th/0212193}{{\tt
  hep-th/0212193}}.

\bibitem{Loewy:2001pq}
A.~Loewy and J.~Sonnenschein, {\it On the holographic duals of n = 1 gauge
  dynamics},  {\em JHEP} {\bf 08} (2001) 007,
  [\href{http://xxx.lanl.gov/abs/hep-th/0103163}{{\tt hep-th/0103163}}].

\bibitem{Sakai:2003wu}
T.~Sakai and J.~Sonnenschein, {\it Probing flavored mesons of confining gauge
  theories by supergravity},  {\em JHEP} {\bf 09} (2003) 047,
  [\href{http://xxx.lanl.gov/abs/hep-th/0305049}{{\tt hep-th/0305049}}].

\end{thebibliography}\endgroup

\end{document}